\documentclass[fleqn,usenatbib]{mnras}

\usepackage{newtxtext,newtxmath}

\usepackage[T1]{fontenc}

\DeclareRobustCommand{\VAN}[3]{#2}
\let\VANthebibliography\thebibliography
\def\thebibliography{\DeclareRobustCommand{\VAN}[3]{##3}\VANthebibliography}

\usepackage{graphicx}	
\usepackage{adjustbox}  
\usepackage{amsmath}	
\usepackage{wasysym}    
\usepackage{bm}         
\usepackage{xspace}
\usepackage{soul}
\makeatletter 
  \patchcmd{\NAT@citex}
    {\@citea\NAT@hyper@{%
      \NAT@nmfmt{\NAT@nm}%
      \hyper@natlinkbreak{\NAT@aysep\NAT@spacechar}{\@citeb\@extra@b@citeb}%
      \NAT@date}}
    {\@citea\NAT@nmfmt{\NAT@nm}%
    \NAT@aysep\NAT@spacechar\NAT@hyper@{\NAT@date}}{}{}

  \patchcmd{\NAT@citex}
    {\@citea\NAT@hyper@{%
      \NAT@nmfmt{\NAT@nm}%
      \hyper@natlinkbreak{\NAT@spacechar\NAT@@open\if*#1*\else#1\NAT@spacechar\fi}%
        {\@citeb\@extra@b@citeb}%
      \NAT@date}}
    {\@citea\NAT@nmfmt{\NAT@nm}%
    \NAT@spacechar\NAT@@open\if*#1*\else#1\NAT@spacechar\fi\NAT@hyper@{\NAT@date}}
    {}{}
\makeatother

\newcommand\Msun{\text{M}_{\astrosun}} 
\newcommand\HI{\ion{H}{I}\xspace} 
\newcommand\HII{\ion{H}{II}\xspace} 
\newcommand\HeI{\ion{He}{I}\xspace} 
\newcommand\HeII{\ion{He}{II}\xspace} 
\newcommand\thesan{\mbox{\textsc{thesan}}\xspace}
\newcommand\thesanone{\mbox{\textsc{thesan-1}}\xspace}
\newcommand\thesantwo{\mbox{\textsc{thesan-2}}\xspace}
\newcommand\thesanwc{\mbox{\textsc{thesan-wc-2}}\xspace}
\newcommand\thesanhigh{\mbox{\textsc{thesan-high-2}}\xspace}
\newcommand\thesanlow{\mbox{\textsc{thesan-low-2}}\xspace}
\newcommand\thesansdao{\mbox{\textsc{thesan-sdao-2}}\xspace}
\newcommand\thesanhr{\mbox{\textsc{thesan-hr}}\xspace}
\newcommand\colt{\mbox{\textsc{colt}}\xspace}
\newcommand\orcid[1]{\href{http://orcid.org/#1}{\adjustbox{trim={-.15\width} {0\height} {-.15\width} {0\height},clip}{\includegraphics[height=10pt]{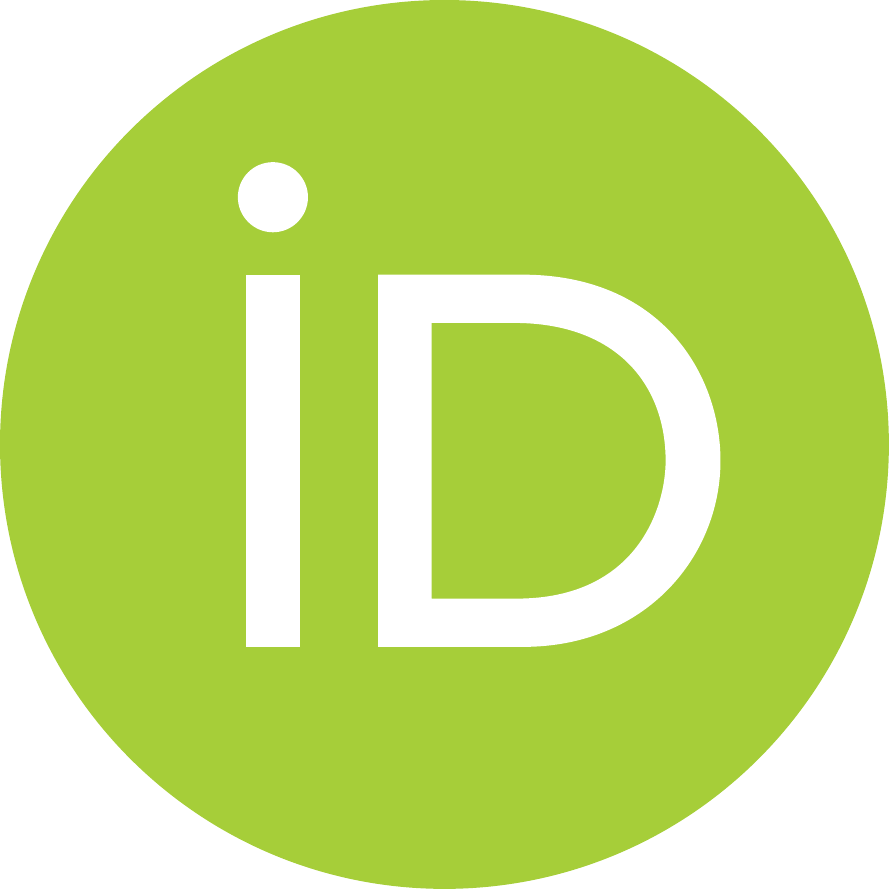}}}}

\title[Escape fractions of reionization-era galaxies]{The \textsc{thesan} project: ionizing escape fractions of reionization-era galaxies}

\author[J.~Y.-C.~Yeh et al.]{%
Jessica~Y.-C.~Yeh\orcid{0000-0002-5721-7679},$^{1}$\thanks{E-mail: \href{mailto:ycyeh@mit.edu}{ycyeh@mit.edu}}
Aaron~Smith\orcid{0000-0002-2838-9033},$^{2,1}$\thanks{E-mail: \href{mailto:aaron.smith@cfa.harvard.edu}{aaron.smith@cfa.harvard.edu}; NHFP Einstein Fellow.}
Rahul~Kannan\orcid{0000-0001-6092-2187},$^{3,2}$
Enrico~Garaldi\orcid{0000-0002-6021-7020},$^{4}$
Mark~Vogelsberger\orcid{0000-0001-8593-7692},$^{1,5}$
\newauthor
Josh~Borrow\orcid{0000-0002-1327-1921},$^{1}$
R\"{u}diger~Pakmor\orcid{0000-0003-3308-2420},$^{4}$
Volker~Springel\orcid{0000-0001-5976-4599}$^{4}$
and Lars~Hernquist$^{2}$
\\%
$^{1}$Department of Physics, Massachusetts Institute of Technology, Cambridge, MA 02139, USA \\%
$^{2}$Center for Astrophysics $\vert$ Harvard $\&$ Smithsonian, 60 Garden Street, Cambridge, MA 02138, USA \\%
$^{3}$Department of Physics and Astronomy, York University, 4700 Keele Street, Toronto, Ontario, Canada, MJ3 1P3 \\%
$^{4}$Max-Planck Institute for Astrophysics, Karl-Schwarzschild-Str.~1, D-85741 Garching, Germany \\%
$^{5}$The NSF AI Institute for Artificial Intelligence and Fundamental Interactions, Massachusetts Institute of Technology, Cambridge MA 02139, USA%
}

\date{Accepted XXX. Received YYY; in original form ZZZ}

\pubyear{2023}

\begin{document}
\label{firstpage}
\pagerange{\pageref{firstpage}--\pageref{lastpage}}
\maketitle

\begin{abstract}
  A fundamental requirement for reionizing the Universe is that a sufficient fraction of the ionizing photons emitted by galaxies successfully escapes into the intergalactic medium. However, due to the scarcity of high-redshift observational data, the sources driving reionization remain uncertain. In this work, we calculate the ionizing escape fractions ($f_\text{esc}$) of reionization-era galaxies from the state-of-the-art \thesan simulations, which combine an accurate radiation-hydrodynamic solver (\textsc{arepo-rt}) with the well-tested IllustrisTNG galaxy formation model to self-consistently simulate both small-scale galaxy physics and large-scale reionization throughout a large patch of the universe ($L_\text{box} = 95.5\,\text{cMpc}$). This allows the formation of numerous massive haloes ($M_\text{halo} \gtrsim 10^{10}\,\Msun$), which are often statistically underrepresented in previous studies but are believed to be important to achieving rapid reionization. We find that low-mass galaxies ($M_\text{stars} \lesssim 10^7\,\Msun$) are the main drivers of reionization above $z \gtrsim 7$, while high-mass galaxies ($M_\text{stars} \gtrsim 10^8\,\Msun$) dominate the escaped ionizing photon budget at lower redshifts. We find a strong dependence of $f_\text{esc}$ on the effective star formation rate (SFR) surface density defined as the SFR per gas mass per escape area, i.e. $\bar{\Sigma}_\text{SFR} = \text{SFR}/M_\text{gas}/R_{200}^2$. The variation in halo escape fractions decreases for higher mass haloes, which can be understood from the more settled galactic structure, SFR stability, and fraction of sightlines within each halo significantly contributing to the escaped flux. Dust is capable of reducing the escape fractions of massive galaxies, but the impact on the global $f_\text{esc}$ depends on the dust model. Finally, active galactic nuclei are unimportant for reionization in \thesan and their escape fractions are lower than stellar ones due to being located near the centres of galaxy gravitational potential wells.
%
\end{abstract}

\begin{keywords}
radiative transfer -- methods: numerical -- galaxies: high-redshift -- cosmology: dark ages, reionization, first stars
\end{keywords}



\section{Introduction}

The formation of the first stars and galaxies marks the end of the cosmic dark ages and the beginning of the Epoch of Reionization \citep[EoR;][]{BrommYoshida2011,LoebFurlanetto2013}. Lyman continuum (LyC; photon energy $h\nu > 13.6\,\text{eV}$) radiation capable of ionizing atomic hydrogen was emitted by a growing number of sources, including young, hot massive stars. These photons primarily interacted with the surrounding interstellar medium (ISM), although a significant fraction successfully traversed through the circumgalactic medium (CGM) of their host galaxies to emerge into the vast intergalactic medium (IGM). The result was to transform the originally cold and neutral atoms into a hot plasma through photoionization and photoheating \citep{Madau1999, Barkana2001}. However, the complex nature of this last major phase transition of the Universe still presents many important unanswered questions. For instance, what are the main sources and sinks of ionizing photons in this period, what determines the morphology and coalescence timeline of ionized bubbles, and how do these translate to observational signatures for current and upcoming facilities? To answer these questions, significant effort has been made to better understand the ionizing source emissivity, either through inferred measurements or theoretical modelling \citep[e.g.][]{Robertson2010,Wise2019,Eide2020}. It is widely believed that star-forming galaxies are the primary sources of reionization \citep{Haardt&Madau2012,Faucher-Giguere2020}. Throughout this paper we focus on the escape fraction $f_\text{esc}$, which we define as the ratio between the number of photons that escape galaxies to reach the IGM and the number of photons intrinsically emitted. As galaxy formation constraints improve, e.g. for stellar population synthesis modelling and stellar-to-halo mass relations, the escape fractions directly dictate the amount of radiation galaxies can contribute to reionization. However, the majority of the measured values of escape fractions to date are not high enough to fully reionize the Universe \citep{Finkelstein2012,Robertson2013}, and thus expose our lack of a complete understanding of reionization.
In the following we briefly summarize observational and theoretical approaches that have been explored to rectify this inconsistency.

There are at least two major challenges when determining $f_\text{esc}$ observationally. First, the high opacity of the intervening neutral IGM at high redshifts means that very limited observational data are available at present. In fact, certain properties such as the intrinsic LyC flux may never be directly observable, while other measurements such as Lyman-$\alpha$ (Ly$\alpha$) absorption spectra become highly saturated. Secondly, it has been shown that $f_\text{esc}$ has large sightline-to-sightline variability \citep{Cen2015}. In other words, individual observations may not be able to yield a representative picture and a large number of galaxies are needed to provide robust constraints. Despite these difficulties, several observations have been made by measuring the leakage of LyC photons, but due to detection biases most of which originate from starburst galaxies at $z \lesssim 3$, and almost all results indicate low $f_\text{esc}$. For example, $f_\text{esc}$ of a few per cent at $z \approx 1$ were detected in star-forming galaxies by \citet{Bridge2010}, \citet{Siana2010}, and \citet{Rutkowski2016}. More optimistically, averages of $f_\text{esc} \sim 10\%$ at $z \approx 3$ were found by \citet{Nestor2013}, \citet{Grazian2017}, \citet{Japelj2017}, \citet{Steidel2018}, and \citet{Pahl2021}.
Recently the Low-Redshift Lyman Continuum Survey (LzLCS) measured a median $f_\text{esc}$ of $4\%$ from galaxies at $z = 0.2$--$0.4$ \citep{Flury2022,SaldanaLopez2022}. Only a few galaxies with uncharacteristically high $f_\text{esc}$ have been reported \citep{Vanzella2018, Izotov2018,Rivera-Thorsen2019}. Looking forward, we expect more data to become available from upcoming facilities, including probing the faint and bright ends of the UV luminosity function with the \textit{James Webb Space Telescope} (\textit{JWST}) and \textit{Nancy Grace Roman Space Telescope} (\textit{NGRST}), as well as 21\,cm cosmology measurements at high-redshift with the Low-Frequency Array (LOFAR), Hydrogen Epoch of Reionization Array (HERA), and Square Kilometer Array (SKA).

The uncertainty in determining escape fractions propagates to different predictions for the global reionization history. In particular, if the dominant photon sources are low-mass galaxies ($M_\text{halo} \lesssim 10^9\,\Msun$), then reionization will have started earlier \citep[e.g.][]{Finkelstein2019}, reaching an IGM ionized volume filling fraction of $0.5$ at $z = 9$. On the other hand, late reionization is favoured if massive galaxies ($M_\text{halo} \gtrsim 10^{10}\,\Msun$) prevail, potentially delaying the midpoint of reionization to around $z = 6.8$ \citep[e.g.][]{Naidu2020}. In this scenario reionization proceeds significantly more rapidly than early reionization models, especially at the tail end \citep{Robertson2015}.
Recently, late reionization has gained popularity, increasing the importance of studying $f_\text{esc}$ from massive galaxies. This is supported by the measurement of a much lower optical depth for electron scattering of cosmic microwave background (CMB) photons than previous studies \citep{Planck2018}, a rapid decrease of the observed Ly$\alpha$ emission \citep{Schenker2014, Mason2019}, and the damping wing absorption on the spectrum of high-redshift quasars \citep{Davies2018}. Collectively, the observations suggest that reionization was nearly complete by $z \approx 6$ \citep{McGreer2015} with large islands of neutral hydrogen remaining in underdense regions below $z \approx 5.5$ \citep{Kulkarni2019}.

The difficulty of obtaining statistically significant data at high redshift ($z \gtrsim 6$) makes theoretical modelling critical in the study of the EoR. However, determining $f_\text{esc}$ with numerical simulations is also an extremely complicated problem since the details can depend on the models used for galaxy formation, feedback recipes, multiphase ISM, and dust physics. Some early studies based on cosmological simulations predict that escape fractions are correlated with the galaxy mass \citep{Gnedin2008, Wise&Cen2009} while some suggest the opposite trend \citep{Yajima2011}. On the contrary, a combination of studies from more recent radiation-hydrodynamic or post-processed radiative transfer simulations indicate a more complicated dependence of $f_\text{esc}$ on halo mass:  $f_\text{esc}$ decreases with halo mass for haloes up to $10^9\, \Msun$  \citep{Paardekooper2015, Xu2016} with high $f_\text{esc}$ in faint galaxies \citep{Kimm2017}. \citet{Ma2020} found the $f_\text{esc}$ increases with halo mass for $M_\text{halo} = 10^8$--$10^{9.5}\,\Msun$ and decreases for more massive haloes at $\gtrsim 10^{11}\,\Msun$. Predicted $f_\text{esc}$ from a wide range of hydrodynamical simulations have been found to be from a few percent to thirty percent \citep{Ma2015,Paardekooper2015, Xu2016,Rosdahl2018,Trebitsch2018,Ma2020,Ma2021,Rosdahl2022}. The potential boost in $f_\text{esc}$ due to binary stars has also been explored \citep{Ma2016, MaFiaschi2022}. However, due to the high computational cost associated with large-volume simulations, most above-mentioned simulations either only have haloes with $M_\text{halo}$ up to $10^9\,\Msun$ or do not have enough statistics in massive haloes ($M_\text{halo} \gtrsim 10^{11}\, \Msun$) and therefore cannot provide a complete picture of the role of more massive galaxies in the reionization process. We note that the recent CoDa simulations \citep{Ocvirk2020} feature large-volume (94.5\,cMpc) boxes and also model haloes down to $\sim 10^8\,\Msun$. The associated ray-tracing analysis by \citet{Lewis2020} shows a decreasing $f_\text{esc}$ with increasing halo mass and highlights the significance of intermediate-mass haloes ($6 \times 10^8 < M_\text{halo}/\Msun < 3 \times 10^{10}$). However, the CoDa\,II\,(III) simulations have relatively low spatial resolution due to the fact that they use a uniform grid code with $23\,(11.5)\,\text{ckpc}$ cells, which impacts both the galaxy formation physics as well as the sourcing and escape of ionizing photons. In fact, \citet{Lewis2020} found that cell escape fractions are bimodal depending on whether the host cell is ionized or neutral. It remains unclear whether coarse resolution within galaxies can accurately capture the physics of escaped emission, and thus it is important to compare different simulations to provide further insights.

In this work, we provide a detailed study of ionizing escape fraction statistics and the underlying properties of the galaxies that contribute to reionization by using the \thesan simulations \citep{Kannan2022,Garaldi2022,Smith2022}. The strength of \thesan is the significantly expanded range of physical processes, based on combining the IllustrisTNG galaxy formation model with a self-consistent on-the-fly treatment of radiation with \textsc{arepo-rt}, while simulating a large volume of the Universe at state-of-the-art resolution. As a result, \thesan successfully produces realistic galaxies that match important physical quantities relevant to the EoR, and has demonstrated its strength with a breadth of topics including 21\,cm cosmology \citep{Kannan2022,Qin2022}, IGM--galaxy connections \citep{Garaldi2022}, Lyman-$\alpha$ emission and transmission \citep{Smith2022,Xu2022}, and line intensity mapping \citep{KannanLIM2022}.

The paper is organized as follows. In Section~\ref{sec:methods}, we describe the \thesan simulations and radiative transfer analysis methodology. In Section~\ref{sec:global_statistics}, we present insights into the global production and escape statistics of LyC photons. In Section~\ref{sec:f_esc_on_halo_properties}, we explore the dependence on halo properties and variation across galaxy populations. In Section~\ref{sec:variations}, we discuss the halo-to-halo variations in escape fractions and characterize the anisotropic behaviour across sightlines from individual haloes. In Section~\ref{sec:caveats}, we discuss several important caveats, including the subresolution modelling, dust pre-absorption of ionizing photons, and contribution of active galactic nuclei (AGNs). Finally, in Section~\ref{sec:summary}, we provide a summary of our main results and possible future investigations.

\begin{figure*}
    \centering
    \includegraphics[width=\textwidth]{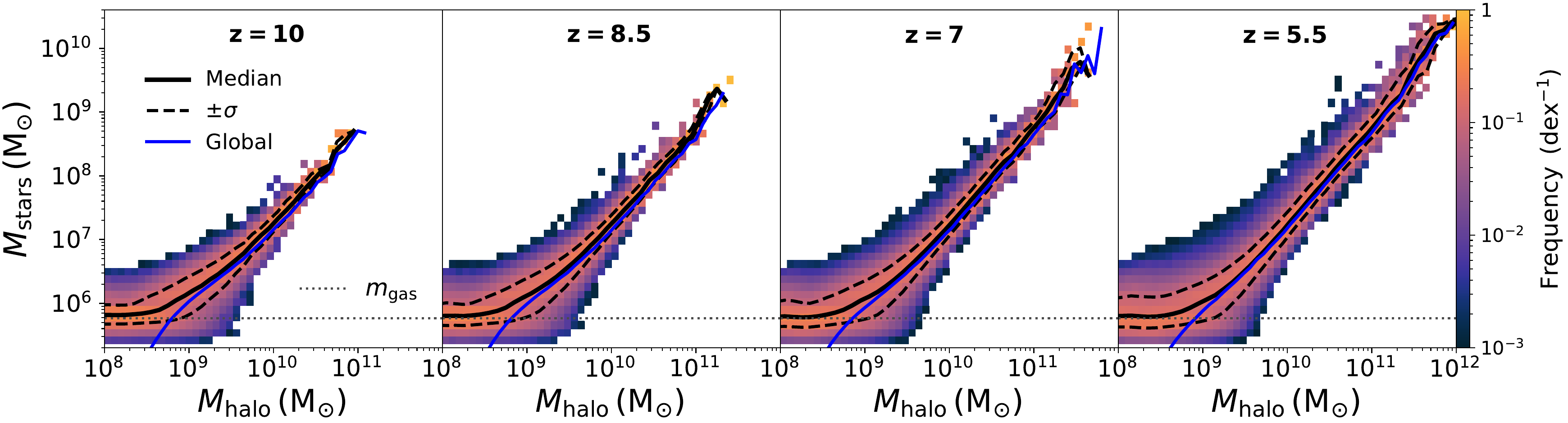}
    \caption{The stellar-to-halo mass relation for all stars within $R_{200}$ for all selected haloes at $z=\{10,8.5,7,5.5\}$. The black solid curves show the median while the black dashed curves show the $16^\text{th}$ and $84^\text{th}$ percentiles of the selected data. The blue curves show the average stellar mass from the full halo catalogue. The mismatch between the selected haloes and the full catalogue comes from the fact that only haloes with at least one star particle are selected. The flattening at the low-mass end is due to the baryonic mass resolution of $m_\text{gas} \approx 5.8 \times 10^{5}\,\Msun$ from the \thesan simulation, as indicated by the gray dotted lines. The coloured histogram illustrates the full distribution with each bin normalized by the number of haloes in the given halo mass range, i.e. the following equation holds: $\int p(M_\text{star}|M_\text{halo})\,\text{d}\log M_\text{star} = 1$.}
    \label{fig:m_star_m200}
\end{figure*}

\section{Methods}
\label{sec:methods}
We briefly describe the \thesan reionization simulations in Section~\ref{sec:thesan}, focusing on aspects of the model that are particularly relevant for our study of escape fractions in the EoR. In Section~\ref{sec:halo_selection} we discuss the halo selection and completeness of our escape fraction catalogues. Finally, in Section~\ref{sec:RT} we summarize the outcomes of the Monte Carlo radiative transfer (MCRT) calculations employed to obtain $f_\text{esc}$ from galaxies extracted from the \thesan simulation.

\subsection{\thesan simulations}
\label{sec:thesan}
The \thesan project is a suite of large-volume radiation-magneto-hydrodynamic simulations that self-consistently model the hydrogen reionization process and the resolved properties of the sources (galaxies and AGN) responsible for it. The simulations were performed with \textsc{arepo-rt} \citep{Kannan2019}, a radiation-hydrodynamic extension of the moving mesh code \textsc{arepo} \citep{Springel2010,Weinberger2020}, which solves the fluid equations on an unstructured Voronoi mesh that is allowed to move along with the fluid flow for an accurate quasi-Lagrangian treatment of cosmological gas flows. Gravitational forces are calculated using a hybrid Tree-PM approach, which splits the force into a short-range force that is computed using an oct-tree algorithm \citep{Barnes1986} and a long-range force, estimated using a particle mesh approach.

Radiation fields are modelled using a moment based approach that solves the zeroth and first moments of the radiative transfer equation \citep{Rybicki1986}, coupled with the M1 closure relation, that approximates the Eddington tensor based on the local properties of each cell \citep{Levermore1984}. They are coupled to the gas via a non-equilibrium thermochemistry module, which self-consistently calculates the ionization states and cooling rates from hydrogen and helium, while also including equilibrium metal cooling and Compton cooling of the CMB \citep[see Section 3.2.1 of][for more details]{Kannan2019}. Both stars and AGN act as sources of radiation, with the spectral energy distribution (SED) of stars taken from the Binary Population and Spectral Synthesis models \citep[BPASS;][]{BPASS2017}. The AGN radiation output is scaled linearly with the mass accretion rate with a radiation conversion efficiency of 0.2 \citep{Weinberger2018} and a \citet{Lusso2015} parametrization for the shape of its spectrum.

The prescriptions for processes happening below the resolution limit of the simulations, such as star and black hole formation and feedback and metal production and enrichment, are taken from the IllustrisTNG model \citep{Weinberger2017,Springel2018,Pillepich2018a,Pillepich2018b,Naiman2018,Marinacci2018,Nelson2018,Nelson2019,Pillepich2019}. The model is augmented with a scalar dust model that tracks the production, growth, and destruction of dust using the formalism outlined in \citet{McKinnon2017}. An additional birth cloud escape fraction parameter, $f_\text{esc}^\text{cloud}=0.37$, is added to mimic the absorption of LyC photons happening below the resolution scale of the simulation. The parameter is tuned such that the simulation reproduces a realistic late-reionization history \citep{Kannan2022}, which matches the observed neutral fraction evolution in the Universe \citep{Greig2017}.

All \thesan simulations follow the evolution of a cubic patch of the universe with linear comoving size \mbox{$L_\text{box} = 95.5$\,cMpc}, and utilize variance-suppressed initial conditions \citep{Angulo2016}. We employ a \citet{Planck2015_cosmo} cosmology (more precisely, the one obtained from their \texttt{TT,TE,EE+lowP+lensing+BAO+JLA+H$_0$} dataset), i.e. $H_0 = 100\,h\,\text{km\,s}^{-1}\text{Mpc}^{-1}$ with $h=0.6774$, $\Omega_\mathrm{m} = 0.3089$, $\Omega_\Lambda = 0.6911$, $\Omega_\mathrm{b} = 0.0486$, $\sigma_8 = 0.8159$, and $n_s = 0.9667$, where all symbols have their usual meanings. In this paper we focus on the highest resolution \thesanone simulation, which has a dark matter mass resolution of $3.12 \times 10^6\,\Msun$ and a baryonic mass resolution of $5.82 \times 10^5\,\Msun$. The gravitational forces are softened on scales of \mbox{$2.2$\,ckpc} with the smallest cell sizes reaching $10$\,pc. This allows us to model atomic cooling haloes throughout the entire simulation volume.

\subsection{Caveats about feedback and ISM physics}
\label{sec:eos}
We now briefly discuss some relevant aspects of the galaxy formation model, which uses a subresolution treatment of the ISM as a two-phase gas where cold clumps are embedded in a smooth, hot phase produced by supernova explosions \citep{Springel&Hernquist2003}. Such subgrid modelling comes with the territory of large-volume simulations with demonstrated agreement with observations down to $z = 0$ \citep{VogelsbergerReview2020}, but certainly has implications for ionizing escape fractions. In particular, the temporary decoupling of wind particles from the hydrodynamics reduces the ability to create low-density channels through which ionizing radiation can escape \citep{Trebitsch2017}. Additionally, the effective equation of state (EoS) parametrization is known to result in too smooth of an ISM. The reduced clumpiness compared to simulations that allow the formation of a cold phase leads to high volume-weighted densities in the ISM. This in turn artificially lowers the median escape fractions along with variations across sightlines and haloes. In the context of the subgrid ISM model used in \thesan, it is also difficult to self-consistently decouple escape fractions from the birth cloud and EoS cells, which correspond to cells with hydrogen number densities larger than $n_\text{H} > 0.13\,\text{cm}^{-3}$. The thermochemistry and opacity equations are treated the same for all cells during RT subcycles between hydrodynamical time-steps, but afterwards cells below the EoS pressure floor are reset to be ionized. Therefore, the birth cloud escape fraction calibration systematically accounts for both unresolved clouds and semitransparent EoS cells. This approach works on larger scales because these uncertainties are bundled into the escape fraction parameter, which is tuned to produce the correct reionization history. We are not aware of any large-volume reionization simulations that claim to fully resolve LyC escape down to the level of individual stars, as this may require subparsec cell sizes. Even simulations like \textsc{sphinx} with a stellar LyC luminosity factor of unity should not be interpreted as properly resolving cloud absorption or escape channels, but rather as a favourable by-product of the combination of resolution, subgrid models, and assumed stellar evolution models, given that the overall escape fractions are not converged with resolution \citep[see Appendix~C of][]{Rosdahl2022}. To help interpret \thesan results more broadly, we explore the quantitative impact of altering the EoS ionization state in post-processing in Appendix~\ref{appendix:EoS}.

There may be several different pathways forward to connect subresolution physics with large-volume galaxy formation simulations. For example, compressible multifluid hydrodynamic schemes can help incorporate multiscale unresolved phase mixing \citep{Weinberger2023} while analytic and tuned prescriptions for cloud-wind interactions can help integrate turbulent mixing and radiative cooling \citep{Fielding2022}. Such effective clumping models may also inform the radiation transport to further our understanding of effective escape fractions through low resolution simulation cells \citep{Buck2022} or via subgrid clumping factors in coarse-grained reionization simulations \citep{Bianco2021}. However, these may still produce too smooth of an ISM for self-consistent escape fractions from galaxy formation simulations. In fact, approximately converged first principles multiphase ISM simulations are only feasible for local stellar neighbourhoods and patches of Milky way disks \citep{tigress,silcc} or low-mass dwarf galaxies \citep{lahen,Steinwandel2020,gutcke}. We remain optimistic that high-resolution multiphase zoom-in and small-volume radiation-hydrodynamic simulations can be sufficiently realistic that comparisons to observations also inform us about feedback and ISM physics \citep[e.g.][]{Hopkins2018, Marinacci2019,Kannan2020}.

\subsection{Halo selection}
\label{sec:halo_selection}
The \thesan simulation uses the friends-of-friends \citep[FoF;][]{Springel2005} algorithm as implemented on-the-fly in \textsc{arepo} to group dark matter particles into resolved structures along with their associated gas and star particles. The SUBFIND algorithm as first described in \citet{Springel2001} is then used to identify gravitationally bound substructures, which form the basis of the haloes hosting the galaxies referred to throughout this paper. However, for clarity we define $M_\text{halo}$ to be the halo mass within the radius at which the average density reaches $200$ times the density of the universe, $M_{200}$. Similarly, we adopt this same definition for $M_\text{stars}$ as the stellar mass within $R_{200}$, the star formation rate (SFR) as the sum of the instantaneous SFRs of gas particles within $R_{200}$, and the virial radius, i.e. $R_\text{vir} \equiv R_{200}$. These definitions are chosen to ensure a consistent accounting of properties contributing to the escape fraction calculations.

At this point we select a subsample of haloes from the full catalogues that contain at least one star and four gas cells within the virial radius. On a practical level, these criteria are chosen to ensure Voronoi mesh reconstructions in ray-tracing are robust and therefore reasonable escape fractions can be defined. Although we do not claim all selected haloes are well resolved, our criteria maximize the completeness in halo selection and thus provide a less biased transition from selected haloes to non-selected haloes in the catalogue. We emphasize that any requirement on the number of gas or star particles induces selection biases, which in the context of the IllustrisTNG galaxy formation model arise from stochastic star formation leading to characteristically bursty star formation histories increasing the variation at the low-mass end of the stellar-to-halo mass relation \citep[e.g.][]{Genel2019,Iyer2020}.

\begin{figure}
    \centering
    \includegraphics[width=\columnwidth]{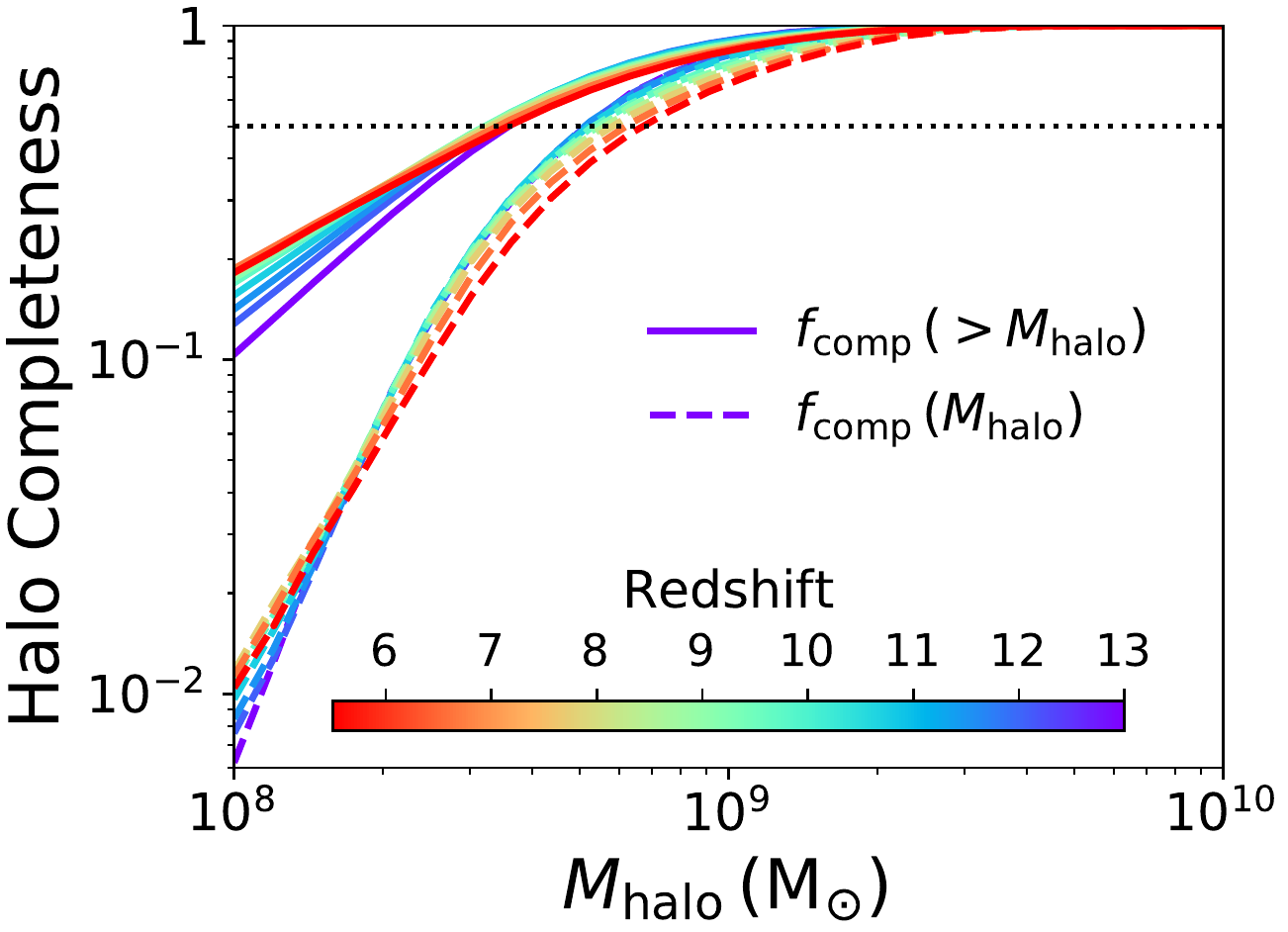}
    \caption{The completeness is defined as the fraction of selected haloes and haloes in the catalogue. The solid curves show the cumulative completeness for all haloes above a given mass and the dashed curves show the completeness at a specific mass at several different redshifts. The cumulative completeness exceeds $90\%$ when $M_\text{halo} \gtrsim 10^{9}\,\Msun$ for nearly all redshifts, and reach $50\%$ when $M_\text{halo} \approx 3 \times 10^8\,\Msun$ as shown by the black dotted line.}
    \label{fig:completion}
\end{figure}

In Fig.~\ref{fig:m_star_m200} we show the stellar-to-halo mass relation for our selected sample of haloes at $z = \{10,8.5,7,5.5\}$. The above selection criteria result in a wide range of haloes from $M_\text{halo} = 10^8$--$10^{12}\,\Msun$. Haloes below $10^8\,\Msun$ are not well-resolved in the \thesan simulation, but these are only expected to be important during the earliest stages of reionization \citep[e.g.][]{Paardekooper2013,Wise2014}. The stellar-to-halo mass relation in \thesan is in agreement with current observational estimates as shown in Fig.~9 of \citet{Kannan2022}. In Fig.~\ref{fig:completion} we show that our selection criteria result in a very high completeness for haloes more massive than $10^9\,\Msun$. That is, our selection covers most of the haloes with $M_\text{halo} > 10^9\, \Msun$ in the catalogue. Specifically, near the midpoint of reionization at $z \approx 7.67$ we achieve greater than $\{10,50,90\}\%$ cumulative completeness at halo masses of $M_\text{halo} \approx \{1.0\times 10^8\,\Msun, 3.2\times 10^8\, \Msun, 1.2\times 10^9\, \Msun\}$.

\subsection{Post-processing radiative transfer}
\label{sec:RT}
As alluded to above, we define the escape fraction of a halo as the ratio between the number of ionizing photons that escape and are intrinsically emitted from all sources within the virial radius, i.e.
\begin{equation} \label{eqn:f_esc}
  f_\text{esc} \equiv \frac{\dot{N}_\text{esc}}{\dot{N}_\text{int}} \, .
\end{equation}
The escape fractions for star particles outside the virial radius of any halo are assumed to be maximal (recall that $f_\text{esc}^\text{cloud} = 0.37$ from the birth cloud) as these directly contribute to reionizing the IGM. The same applies for stars in haloes with fewer than four gas particles (non-selected haloes), which do not possess enough opacity to lead to a significant amount of photon absorption in agreement with calculations of $f_\text{esc}$ in low-mass haloes.
The intrinsic photon rate for each subhalo (including stars outside $R_{200}$) is provided in the Ly$\alpha$ emission catalogue described in \citet{Smith2022}, which is calculated from the Binary Population and Spectral Synthesis models \citep[BPASS version 2.2.1;][]{BPASS2017} with a Chabrier IMF \citep{Chabrier2003}. The photon rates within $R_{200}$ are calculated from the same exact age--metallicity tables as part of the radiative transfer calculations within the Cosmic Ly$\alpha$ Transfer (\colt) code \citep{Smith2015, Smith2019}. This ensures consistent results for global properties when accounting for all ionizing sources. The details of the photoionization physics and MCRT implementation are described in \citet{smith2022disk}, and we only summarize the main features and differences here. Most importantly, we perform passive RT as our goal is to obtain escape fraction statistics based on the on-the-fly ionization states in the \thesan simulation rather than recompute new ones based on the assumption of photoionization equilibrium. Our escape fraction calculations employ $10^5$ photon packets for haloes with fewer than $10$ star particles, while the photon count increases linearly with the number of star particles up to a maximum of $10^7$ packets for well-converged statistics. Similar to \thesan, we model the multifrequency ionizing radiation field in three bands, \HI, \HeI, and \HeII with energy edges of $[13.6, 24.59, 54.42, \infty)$\,eV. We employ continuous absorption algorithms for the treatment of photoionization, with frequency-dependent cross-sections taken from \citet{Verner1996}. The emission direction is isotropic and the initial position corresponds to the location of the star. We employ native ray-tracing through the Voronoi tessellation.

Dust absorption has a non-negligible effect on the escape of LyC photons and has been discussed in previous studies \citep[see e.g.][]{Puglisi2016,Tacchella2018,Ma2020}. \thesan integrates the production and destruction of dust based on the empirical model of \citet{McKinnon2017}, and treats dust as a property of the gas resolution elements. As in \citet{smith2022disk}, we can isolate the pre-absorption of ionizing photons by dust, according to
\begin{equation} \label{eqn:f_esc_species}
  f_\text{abs} \equiv \int_0^\ell k_\text{a,d} \, e^{-k_\text{a} \ell'} \text{d}\ell' \, ,
\end{equation}
where $k_\text{a,d}$ and $k_\text{a}$ denote the dust and total absorption coefficients along the path integration of length $\ell$. The absorption by hydrogen and helium can also be calculated in a similar way by replacing $k_\text{a,d}$ with $k_\text{i} = n_\text{i}\sigma_\text{i},\ i\in \{\HI, \HeI, \HeII\}$. We note that it is challenging to resolve dust in ionized \HII regions with the resolution of the \thesan simulation because of the ISM and feedback models. Hence, the fiducial MCRT calculation in this paper assumes a no-dust scenario, and we postpone a more detailed estimation of the effect of dust absorption to Sec.~\ref{sec:caveats}. Finally, the ray-tracing procedure also allows us to conveniently track the average distance to absorption, defined as:
\begin{equation} \label{eqn:mean_dist}
  \langle \ell \rangle \equiv \ell^{-1} \int_0^\ell \ell' e^{-k_\text{a} \ell'} \text{d}\ell' \, ,
\end{equation}
which provides complementary information about the ionization and dust absorption of photons that we explore in later sections.

\begin{figure*}
    \centering
    \includegraphics[width=\textwidth]{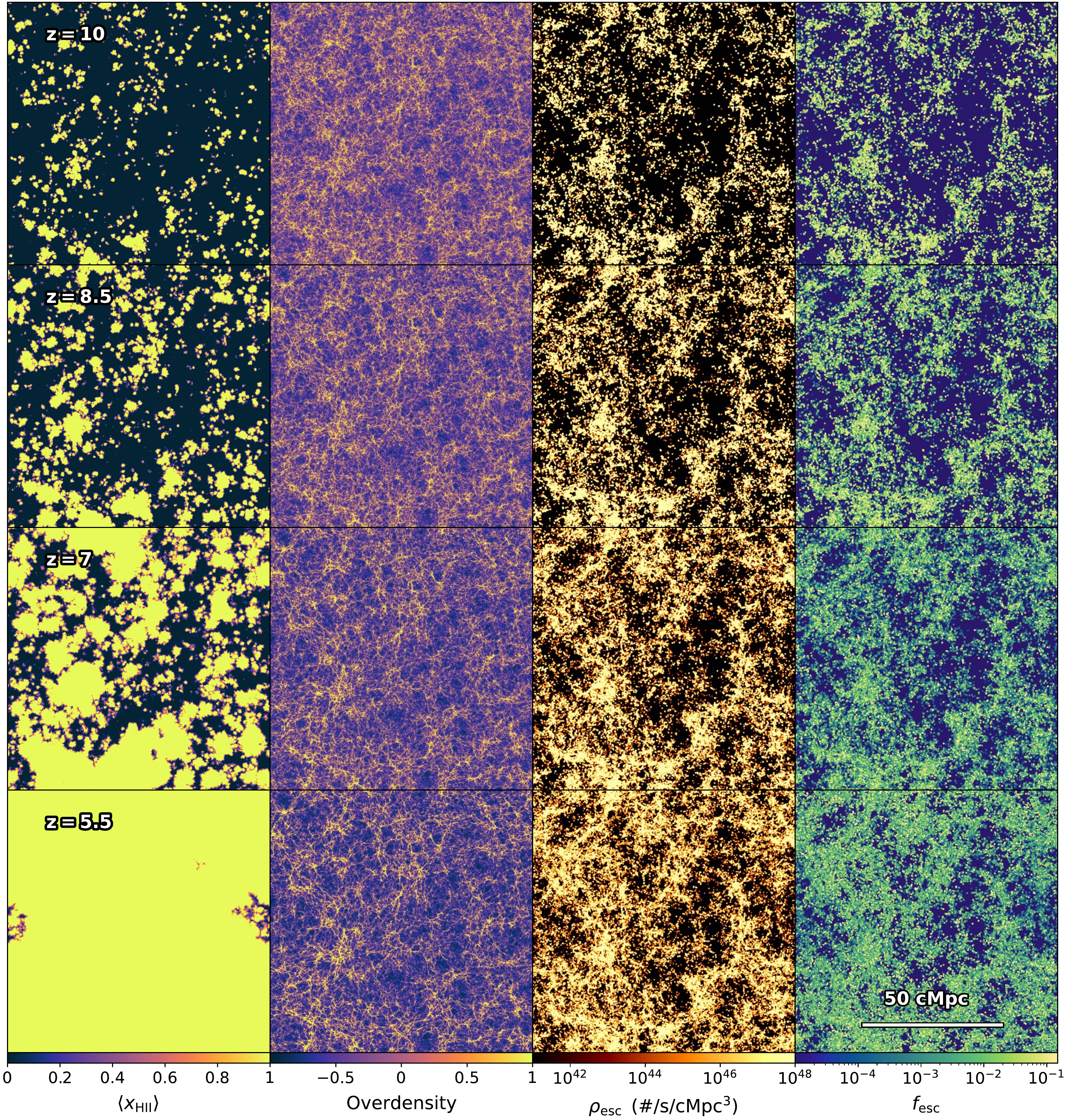}
    \caption{\textit{Left to right:} Visualization of the spatial distribution of the \HII fraction $\langle x_{\HII} \rangle$, gas overdensity $\delta \equiv \rho / \bar{\rho} - 1$, escaped ionizing photon rate densities $\rho_\text{esc}$, and LyC escape fractions $f_\text{esc}$, covering the same $95.5\times95.5\times 9.55\, \text{cMpc}^3$ subvolume at redshifts $z = \{10,8.5,7,5.5\}$, from top to bottom. The similarity in morphology among these quantities is striking, and indicates that environment also influences the production and escape of ionizing photons in the EoR. }
    \label{fig:proj}
\end{figure*}

\begin{figure*}
    \centering
    \includegraphics[width=\textwidth]{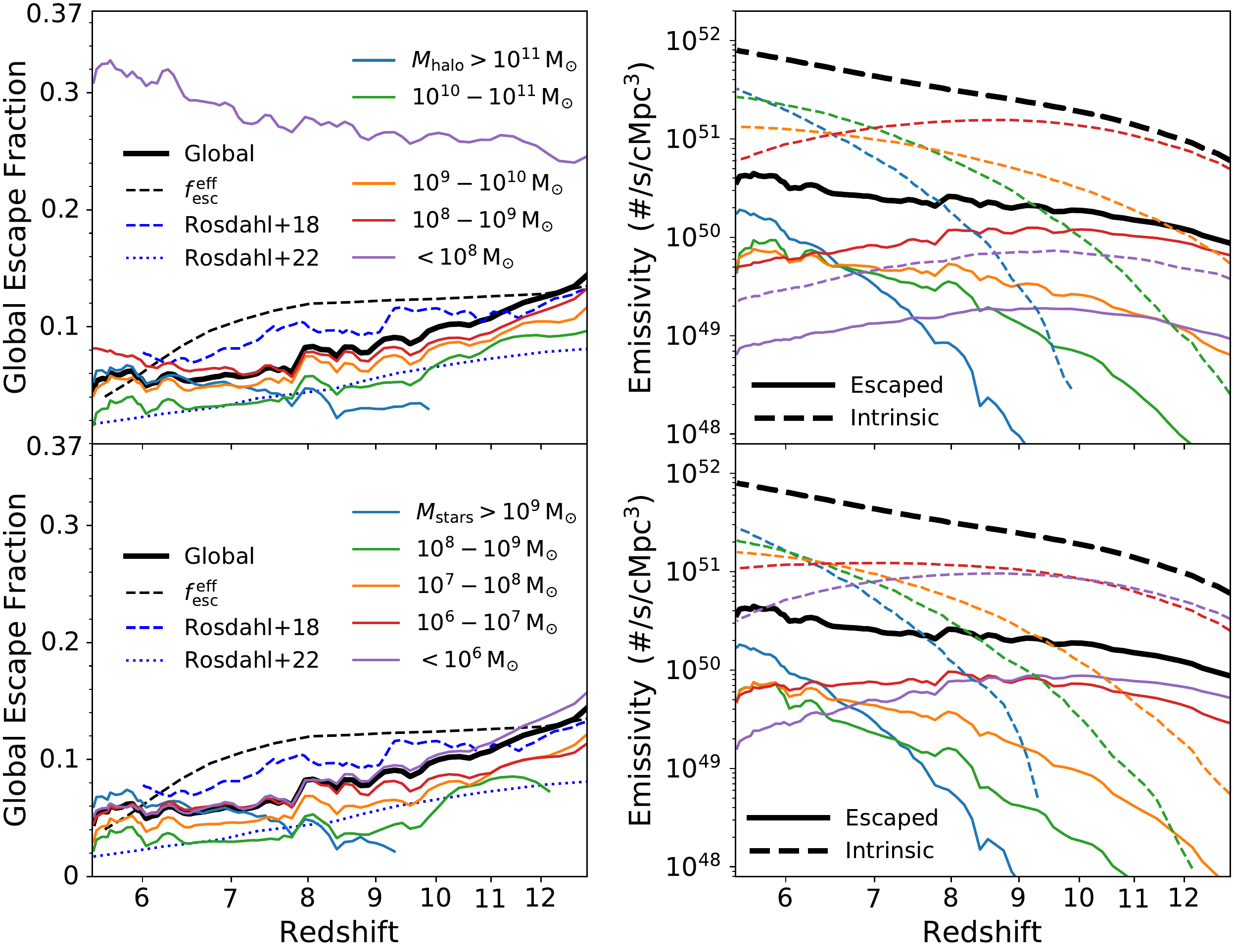}
    \caption{\textit{Left:} Global photon rate weighted escape fractions $\langle f_\text{esc} \rangle \equiv \sum \dot{N}_\text{esc} / \sum \dot{N}_\text{int}$ as a function of redshift (thick black curves). This contains all photon sources in the simulated volume, including all resolved haloes and a small fraction of unassigned star particles. The overall escaped photon rates are obtained using the $f_\text{esc}$ calculated by \colt for all selected haloes and assuming maximal escape $f_\text{esc}^\text{cloud} = 0.37$ for the remaining sources. We also show global escape fractions for haloes in five different halo (upper panels) and stellar (lower panels) mass ranges (coloured curves), which reflect the nontrivial mass dependence with redshift. The lowest mass haloes have lower completeness for stars within the virial radius and thus have maximal escape properties. Overall, the weighted-$f_\text{esc}$ decreases with increasing halo mass except for the most massive haloes (either $M_\text{halo} > 10^{11}\,\Msun$ or $M_\text{stars} > 10^9\,\Msun$). The black dashed curve shows the effective escape fraction $f_\text{esc}^\text{eff}$ calculated as described in \citet{Kannan2022}. For reference, we also include the result from \citet[][blue dashed curves]{Rosdahl2018} and \citet[][blue dotted curves]{Rosdahl2022}. \textit{Right}: Global intrinsic (solid) and escaped (dashed) photon rate densities as functions of redshift (thick black curves), also plotted with the same mass bin ranges (coloured curves). The emissivity evolution is mainly due to the formation of massive galaxies at lower redshifts, with a slightly flattened slope when going from intrinsic to escaped values. The halo and stellar mass dependence clearly show that massive haloes become the dominant photon sources by $z \approx 7$.}
    \label{fig:global_more}
\end{figure*}

\section{Global statistics}
\label{sec:global_statistics}

\subsection{Large-scale structure}
In Fig.~\ref{fig:proj}, we present a visualization of the spatial distribution of the volume-weighted \HII fraction $\langle x_{\HII} \rangle$, gas overdensity $\delta \equiv \rho / \bar{\rho} - 1$, escaped ionizing photon rate densities $\rho_\text{esc}$, and LyC escape fractions $f_\text{esc}$ across a $95.5 \times 95.5\times 9.55 \,\text{cMpc}^3$ subvolume at redshifts $z=\{10,8.5,7, 5.5\}$. $\rho_\text{esc}$ and $f_\text{esc}$ are calculated by dividing the subvolume into a grid and summing up the contributing haloes in each pixel region. We also apply a Gaussian filter over the grid to smooth out the halo distribution. The similarity in morphology among the four quantities can be readily seen. At $z \gtrsim 8$, the reionization of hydrogen corresponds with the highest overdensity regions, which are the earliest sites of galaxy formation and are thus capable of providing the most intrinsic and escaped ionizing photons at these epochs. The \HII regions then expand and fill the entire simulation volume at the end of the simulation at $z = 5.5$ (bottom panel). Regions with high $\rho_\text{esc}$ and $f_\text{esc}$ also generally align with the high overdensity regions at higher redshifts, and they expand into the nearby regions as time increases. At the end, even regions with overdensity less than zero (underdensities) can have relatively high $\rho_\text{esc}$ and $f_\text{esc}$. This emphasizes the important role environment plays during reionization, and neighborhoods without significant star formation can be externally reionized via the ionizing sources nearby.

\subsection{Redshift evolution}
We next present quantitative results for the global statistics, which are critical for understanding the evolution of the IGM ionized volume fraction and therefore the reionization history \citep{Madau1999}. In the left-hand panels of Fig.~\ref{fig:global_more} we show the global escape fraction (thick black curves) defined as $\langle f_\text{esc} \rangle \equiv \sum \dot{N}_\text{esc} / \sum \dot{N}_\text{int}$, where the summations are over all photon sources. In particular, this includes all resolved subhaloes with MCRT calculations but also all stars outside the virial radius of any halo assuming maximal escape fractions. In fact, the plotted range corresponds to $f_\text{esc}^\text{cloud} = 0.37$ for consistency with the birth cloud value used in \thesan to match available constraints on the observed reionization history. To reduce distracting numerical fluctuations, we have applied Savitzky-Golay smoothing filter to all values. To show the dependence on the halo mass, in the upper panels we also calculate the global contributions from sources split into five different mass ranges defined by the following thresholds: $M_\text{halo} \in (0, 10^8, 10^9, 10^{10}, 10^{11}, \infty)\,\Msun$. Likewise, in the lower panels we separate the dependence on the stellar mass in the following ranges: $M_\text{stars} \in (0, 10^6, 10^7, 10^8, 10^9, \infty)\,\Msun$. In comparison to previous similar studies \citep[e.g.][]{Paardekooper2015, Rosdahl2018}, our larger simulation volume provides us with significantly more data for massive galaxies. For reference, we also include the result from \citet{Rosdahl2018} as shown with the blue-dashed curve (the luminosity-weighted mean over the last 100\,Myr $f_{\text{esc},100}$ from the binary SED model in their Fig.~13), and \citet{Rosdahl2022} with the blue-dotted curve ($f_{\text{esc},100}$ in their Fig.~16). Last, we compare our post-processing escape fractions to the effective escape fractions $f_\text{esc}^\text{eff}$ (black-dashed lines) calculated in Sec. 3.3 of \citet{Kannan2022} using the equations outlined in \citet{Madau2017}. Our results show that the effective escape fractions $f_\text{esc}^\text{eff}$ overestimate the escape fractions before $z\approx 6$. This is due to the oversimplification of the equation used to obtain $f_\text{esc}^\text{eff}$, which only treats the LyC production and absorption within a clumpy medium and overlooks smaller structures. Thus, a more careful approach is necessary to determine escape fractions from simulations. In addition, our ray-tracing RT approach results in a redshift-dependent, galaxy-mass dependent $f_\text{esc}$, which is not seen in the zeroth-order estimation of $f_\text{esc}^\text{eff}$. 

In the right-hand panels, we show the total intrinsic (dashed curves) and escaped (solid curves) cosmic photon rate densities, i.e. normalized by the volume of the \thesan simulation. The intrinsic photon rate density rises from $\sim 10^{51}\,\text{photons~s}^{-1}\,\text{cMpc}^{-3}$ at $z = 13$ to $\sim 10^{52}\,\text{photons~s}^{-1}\,\text{cMpc}^{-3}$ at $z = 5.5$, while the escaped photon rate density has a slightly flatter evolution over the same period of time. In fact, there is a clear decrease in the global effective escape fraction from $\approx 15\%$ to $\approx 5\%$, with varying behaviour across the individual halo and stellar mass ranges. Notably, in our model the escape fractions of the lowest mass unresolved haloes ($<10^8\,\Msun$) are close to the maximal values but have a small contribution to reionization due to their low intrinsic photon rates. As reionization progresses the emissivity transitions from being dominated by atomic cooling haloes ($M_\text{halo} \approx 10^8$--$10^9\,\Msun$) to more massive haloes. The suppressed star formation in this mass range is likely due to photoheating feedback by the reionization process itself, which reduces the accretion of gas onto smaller haloes \citep[e.g.][]{Gnedin2014, Dawoodbhoy2018, Wu2019a}.

\begin{figure}
    \centering
    \includegraphics[width=\columnwidth]{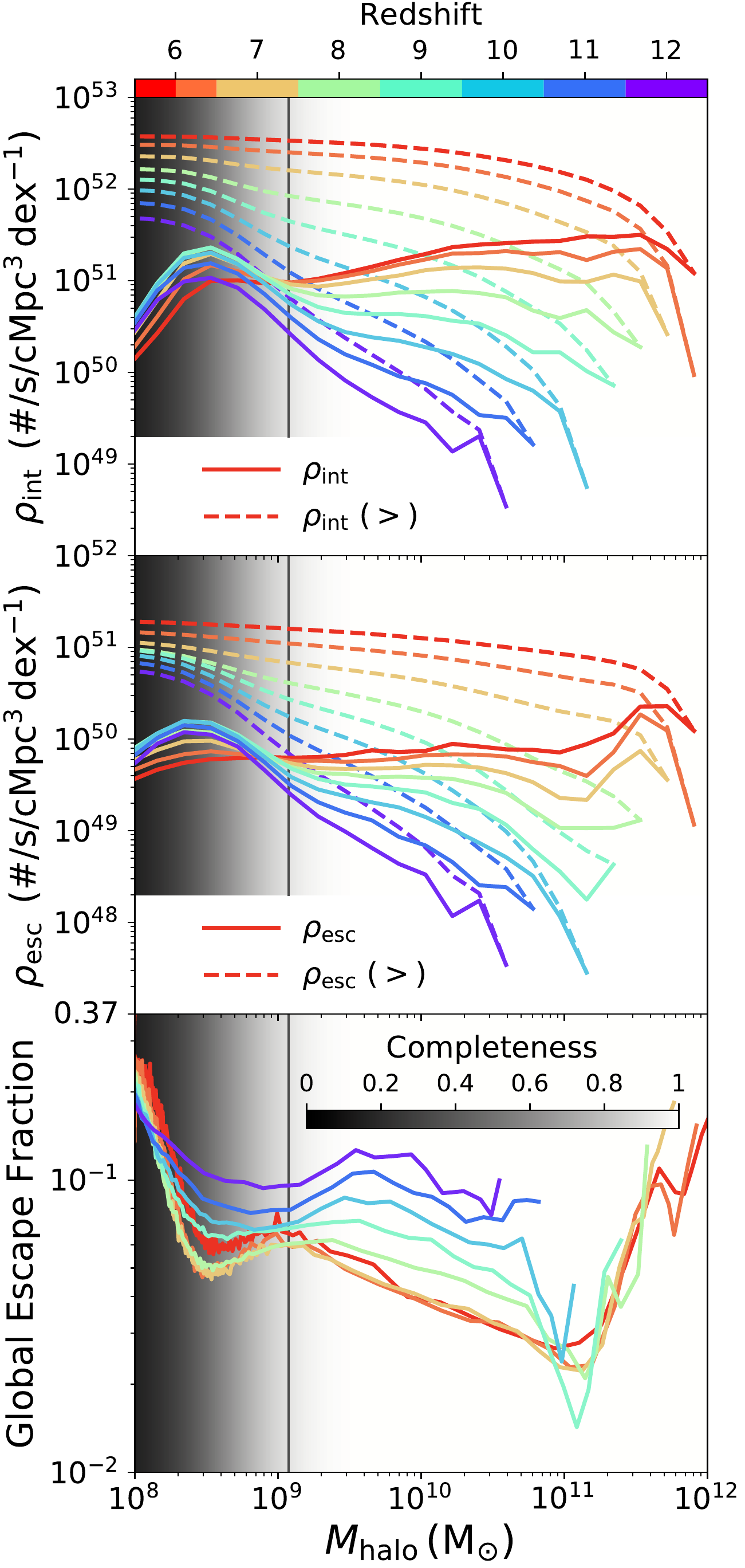}
    \caption{The distribution of global intrinsic (top) and escaped (middle) photon rate densities along with the escape fraction (bottom) as functions of halo mass for several redshift ranges covering $z = 5.5$--$13$ as indicated in the colour bar. The solid curves are mass distributions while the dashed curves are cumulative values from the massive end. The grayscale background shows the cumulative completeness of the halo selection, and the solid black line indicates the point of $90\%$ completeness. Haloes at the low-mass end ($M_\text{halo} \lesssim 10^9\,\Msun$) are less resolved while more massive ones ($M_\text{halo} \gtrsim 10^{11}\,\Msun$) only form at lower redshifts given the moderate volume of the \thesan simulation. Lower mass haloes are the main intrinsic and escaped photon sources at $z \gtrsim 8$ while higher mass haloes become dominant at lower redshifts. The most massive haloes are especially strong sources of escaped photons at $z \lesssim 7$ as indicated by the noticeable peak in the middle panel and upturn in the escape fraction shown in the bottom panel.}
    \label{fig:M_pdf_all}
\end{figure}

\begin{figure}
    \centering
    \includegraphics[width=\columnwidth]{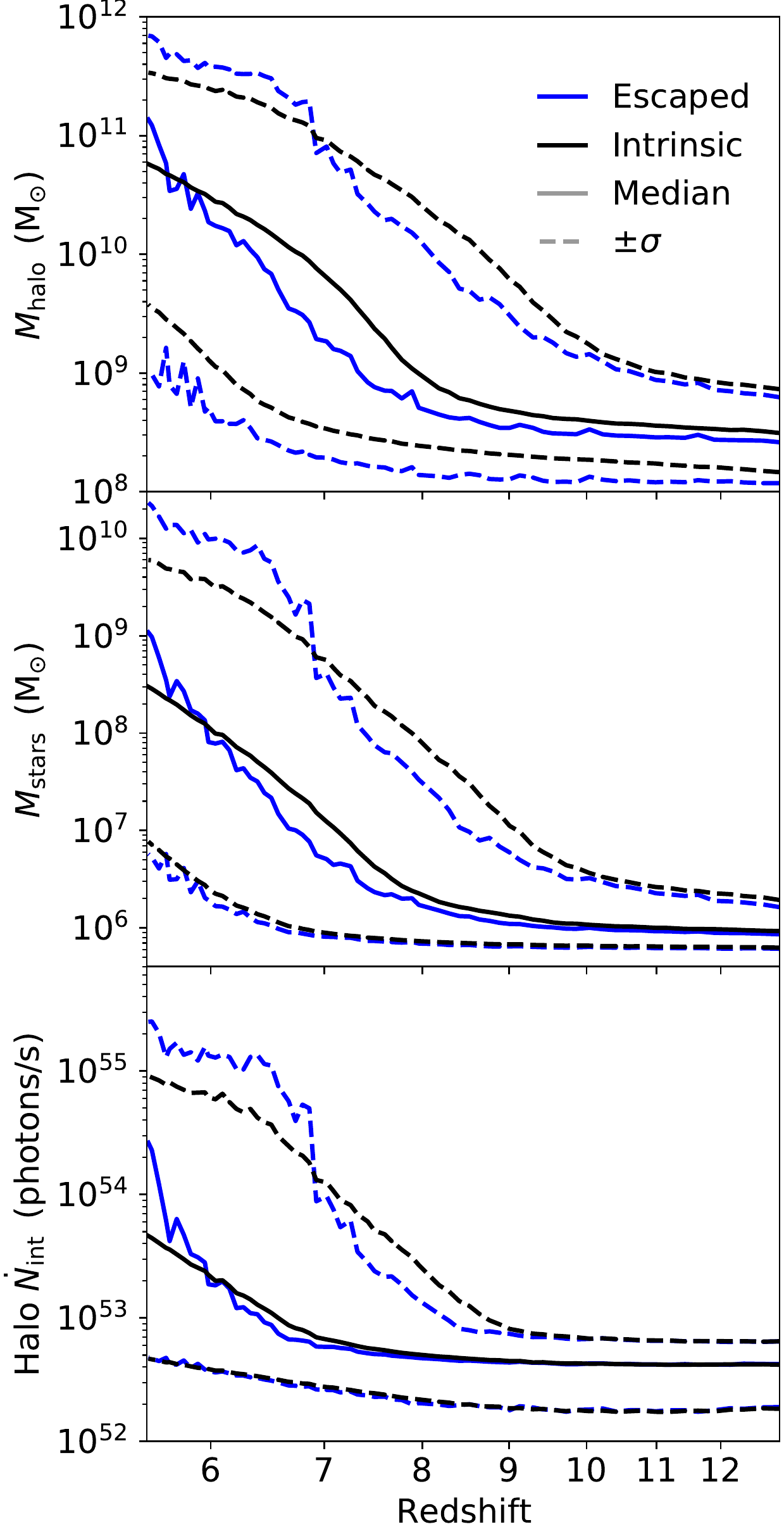}
    \caption{Halo masses, stellar masses, and halo intrinsic photon rates (top, middle, and bottom panels, respectively) that contribute to $\{16,50,84\}\%$ of the global escaped and intrinsic photons (blue and black curves, respectively) when sorted by these quantities across redshifts $z = 5.5$--$13$. The increasing median $M_\text{halo}$, $M_\text{stars}$, and $\dot{N}_\text{int}$ as redshift decreases clearly illustrates the growing importance of massive and bright haloes towards lower redshifts.}
    \label{fig:weighted_mhalo}
\end{figure}

\begin{figure}
    \centering
    \includegraphics[width=\columnwidth]{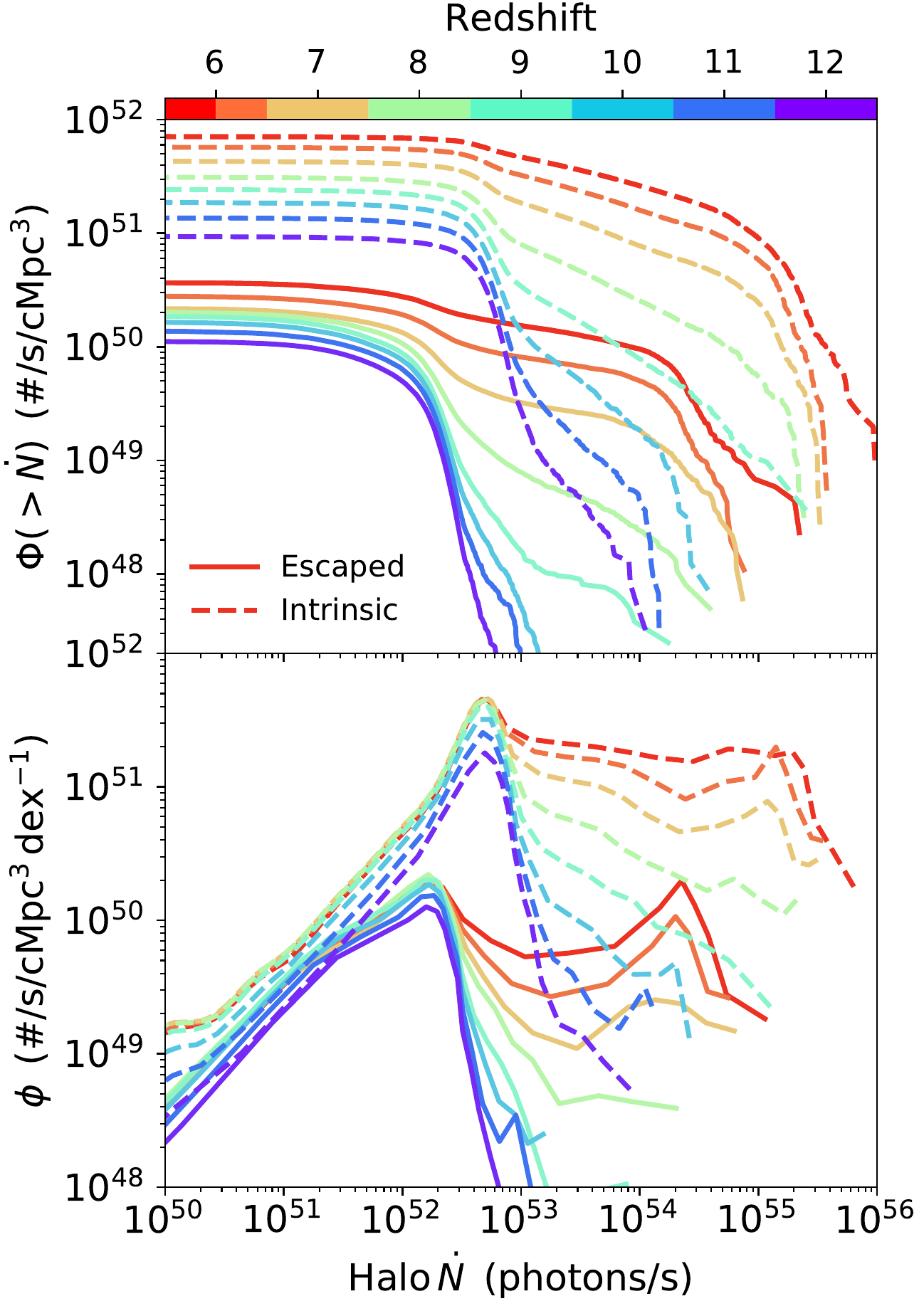}
    \caption{\textit{Top:} Cumulative photon rate density functions for the intrinsic (dashed) and escaped (solid) photon rates $\dot{N}$ as a function of halo luminosity for all haloes in redshift bins covering redshifts from $z = 5.5$--$13$ as ranges indicated in the colour bar. There is a clear evolution revealing an increase of bright haloes starting to dominate the total photon rate density as redshift decreases. The rightmost points show the largest halo photon rate in the redshift bin, which also increases with the formation of more massive haloes. The curves flatten off when the cumulative contribution of fainter haloes is no longer significant. \textit{Bottom:} Distributions of photon rate densities for both intrinsic and escaped halo photon rates. The combined contribution of massive haloes gains importance as redshift decreases in both escaped and intrinsic emission. The peaks at the low-mass end again indicate resolution limits and the point below which fainter haloes do not play important roles.}
    \label{fig:ndot_cdf}
\end{figure}

\begin{figure*}
    \includegraphics[width=\textwidth]{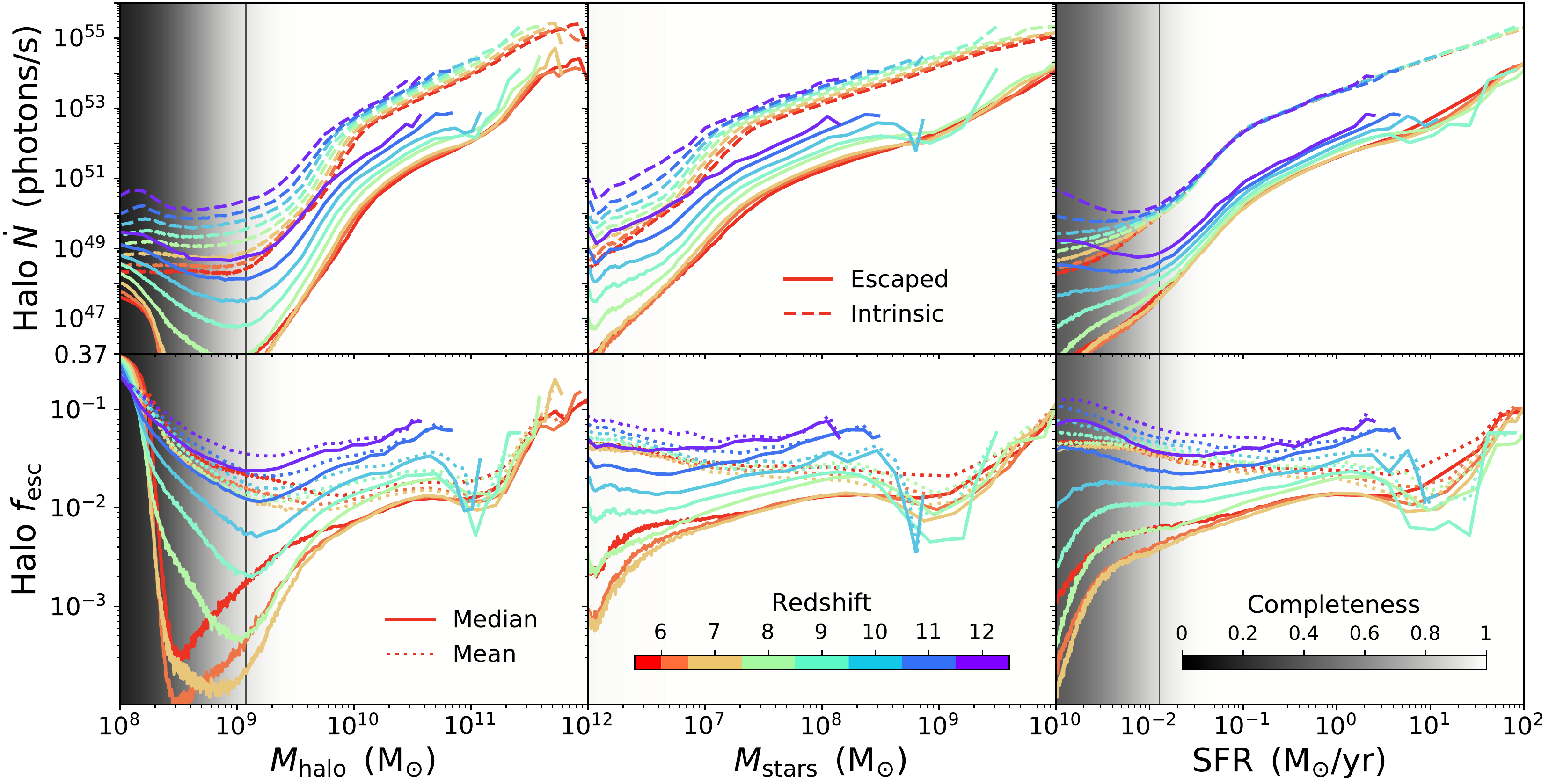}
    \caption{\textit{Top:} The median values of the intrinsic (dashed) and escaped (solid) photon rates as a function of halo mass (left), stellar mass (middle), and SFR (right) for all selected haloes across several redshift ranges covering $z = 5.5$--$13$ as indicated in the colour bar. Haloes falling within the same redshift ranges are combined. The binning is adapted based on the halo number distribution with more bins towards the high-mass end to account for the fewer amount of haloes available in these ranges. The figure includes all the selected galaxies with criteria defined in \ref{sec:halo_selection}, ranging from $10^8$--$10^{12}\,\Msun$. The grayscale background shows the cumulative completeness of the halo selection, and the solid black line indicates the point of $90\%$ completeness. The intrinsic photon rate increases as halo mass increases, and the flattening at the low-mass end is again due to the baryonic mass resolution of the \thesan simulation. The escaped photon rate drops to the minimal at $M_\text{halo} \approx 10^{9}\,\Msun$ and increases beyond this point. \textit{Bottom:} The median (solid) and mean (dashed) of $f_\text{esc}$ as a function of $M_\text{halo}$ with the same binning method and halo selection criteria as above. The escape fractions reach the maximal value of  $0.37$ for lowest mass haloes, and decrease as a function of halo mass afterwards. $f_\text{esc}$ reach the minimal values at $M_\text{halo} \approx 10^9\,\Msun$. For haloes more massive than this value, the $f_\text{esc}$ gradually increases as halo mass increases. Some very massive haloes at low redshifts ($M_\text{halo} > 10^{11}\,\Msun$) can reach $f_\text{esc}=0.1$. A clear redshift dependence can be seen from the figure with higher $f_\text{esc}$ at higher redshifts.}
    \label{fig:f_esc_no_dust}
\end{figure*}

\subsection{Importance of halo mass}
The increasing photon rate densities can also be understood by considering the redshift evolution of the halo and stellar mass distributions towards more massive objects. Conceptually, the intrinsic luminosity of a galaxy is roughly proportional to its SFR, which is correlated with the halo and stellar mass. In Fig.~\ref{fig:M_pdf_all} we show the global intrinsic (top) and escaped (middle) photon rate density distributions along with the escape fraction (bottom) as functions of halo mass ($10^8$--$10^{12}\,\Msun$) for several different redshift bins across the range $z = 5.5$--$13$ as indicated in the colour bar. Data in the same redshift bins are combined to improve the statistics. The solid curves are the underlying mass distributions while the dashed curves represent cumulative values from the massive end. For reference, we also include a grayscale background to show the cumulative completeness of the halo selection. The solid black line indicates the point of $90\%$ completeness, which serves as a reasonable signal for the resolution limit of our escape fraction analysis. At nearly all redshifts the low-mass peak coincides with the range of marginally resolved haloes in the simulation $M_\text{halo} \approx 10^8$--$10^9\,\Msun$, i.e. haloes below this have low stellar occupation fractions so do not emit significant ionizing radiation. The low-mass bump gradually becomes less dominant with time as more massive haloes start to dominate the production budget. Furthermore, the increasing intrinsic photon rate densities from the formation of more massive galaxies changes the negative slope at $z\approx 7.5$--$12.5$ to the positive slope at $z<7.5$, with the exception at $M_\text{halo} \gtrsim 3\times10^{11}\,\Msun$ where the slope drives downward due to the rapidly decreasing number of haloes in this range given the limited simulation volume. This shows that the combined photon rate densities from massive haloes eventually exceed the contribution from lower mass haloes. We emphasize that the intrinsic LyC production is a result of the galaxy formation model, and smaller simulation boxes may significantly underrepresent the role of bright galaxies. A similar trend is also identified in the distribution of escaped photon rate densities (middle panel) with a notable difference at lower redshifts, where the small bumps at $M_\text{halo} > 10^{11}\,\Msun$ are caused by the sharply increasing escape fractions at the massive end in our model. In bottom panel, we observe decreasing global escape fractions as the halo mass increases up until $M_\text{halo} \sim 10^{11}\,\Msun$. \citet{Lewis2020} post-processed Coda~II simulations and also reported a decreasing escape fraction with halo mass. We note that the uptick at $M_\text{halo} > 10^{11}\,\Msun$, which is a unique feature from our post-processing results, can likely be attributed to the different feedback mechanisms in the simulations. For example, AGN sources are ignored in the CoDa simulations even though they are known to impact star formation in high-mass galaxies. In addition, even if the direct contribution of AGN to escaped emission is negligible, they can still alter the thermal and ionization states within their host galaxies.

We directly demonstrate the importance of massive galaxies in Fig.~\ref{fig:weighted_mhalo}, where we plot the corresponding $M_\text{halo}$ (top panel), $M_\text{stars}$ (middle panel), and halo $\dot{N}_\text{int}$ (bottom panel) that contribute to $\{16,50,84\}\%$ of the global escaped (blue) and intrinsic (black) photons when sorted by these quantities, all as functions of redshift. The median $M_\text{halo}$, $M_\text{stars}$, and $\dot{N}_\text{int}$ increase over time in terms of both intrinsic and escaped photons. Moreover, a particularly sharp increase in the escaped distribution can be seen below $z \lesssim 7$, emphasizing the role of the brightest and most massive galaxies in accelerating reionization. Their importance mainly stems from their exceedingly high intrinsic photon rates in combination with moderately higher escape fractions ($f_\text{esc} \sim 0.1$). Similarly, although there are larger numbers of faint galaxies with high $f_\text{esc}$, their combined impact on reionization is minor. These conclusions are consistent with the late-reionization model supported by recent observations and simulations \citep[e.g.][]{Naidu2020,Ma2020,Yung2020}.

\section{Dependence on halo properties}
\label{sec:f_esc_on_halo_properties}

\subsection{Photon rate density functions}
We now investigate the key physical quantities that determine the escape fractions of individual haloes. The dependence on halo mass has been studied in previous cosmological simulations where the results suggest an anticorrelation between halo mass and escape fraction \citep{Paardekooper2015,Kimm2017}. However, these specific studies were limited to haloes with $M_\text{halo} \lesssim 10^9 \Msun$ due to the high computational cost associated with simulating larger volumes at comparable resolutions. Here we take advantage of the significantly improved statistics provided by the \thesan simulation to explore the halo mass dependence of $f_\text{esc}$ extended to higher mass galaxies. Specifically, we examine trends for individual haloes covering a mass range of $M_\text{halo} \in 10^8$--$10^{12}\,\Msun$. Haloes with masses below this range are generally not resolved in the simulation. On the other hand, haloes with masses above this range are not found in the \thesan simulation due to the limited volume of the simulation.

To characterize the halo source population, in the top panel of Fig.~\ref{fig:ndot_cdf} we show cumulative photon rate density against the halo photon rate for the intrinsic (dashed) and escaped (solid) photon rates $\dot{N}$ for all haloes in the same redshift bin ranges as before covering $z = 5.5$--$13$. The photon rate density in this section includes only stars in selected haloes within their virial radii. This is different from Fig.~\ref{fig:M_pdf_all}, where all LyC sources are included. There is a clear evolution with redshift due to the increase of bright haloes as they start to dominate the total photon rate density. The rightmost ends of the curves show the largest photon rate from a single halo at a given redshift, which also increases with time as more massive haloes form. The curves become increasingly flat as fainter halo sources no longer contribute significantly compared to brighter ones. The bottom panel of Fig.~\ref{fig:ndot_cdf} shows the distribution of photon rate densities, again with intrinsic (dashed) and escaped (solid) photon rates in the same redshift range $z = 5.5$--$13$. The shape of the escaped photon rate density distribution resembles the intrinsic one with overall lower values and more evident peaks at the lower redshifts. This is due to the contribution from massive haloes with moderate $f_\text{esc}$ that are formed at low redshifts.

\begin{figure*}
    \centering
    \includegraphics[width=\textwidth]{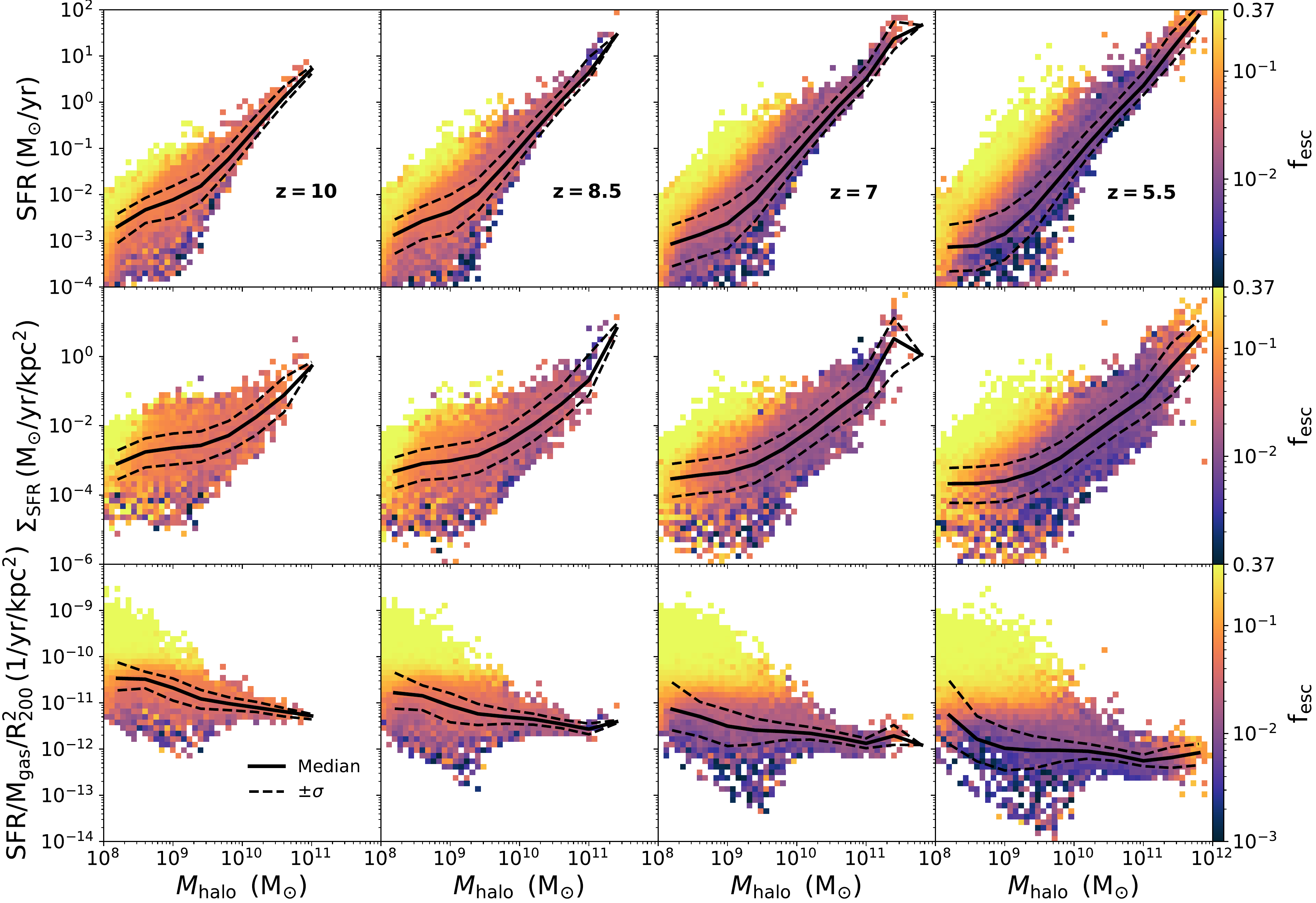}
    \caption{Distribution of haloes at $z=\{10,8.5,7,5.5\}$, from left to right, with various masses and SFRs (top), surface SFR ($\Sigma_\text{SFR}$ as defined in \citet{Pahl2022} with $r_e=$ half stellar mass radius, middle), and modified surface SFR per gas mass ($\overline{\Sigma}_\text{SFR} = \text{SFR}/M_\text{gas}/R_{\rm 200}^2$, bottom). The 2D histograms are coloured by mean $f_\text{esc}$ with $50$ bins in both axes, and each bin is normalized by the number of haloes as in Fig.~\ref{fig:m_star_m200}. The median values are show in the black solid curves and the $16^{\rm th}$ and $84^{\rm th}$ percentiles are shown in the black dashed curve. At each redshift, haloes with relatively low masses and high SFRs have the highest escape fractions. For massive haloes ($M_\text{halo} > 10^{10}\,\Msun$), $f_\text{esc} < 10\%$ even with high SFRs. In addition, in the low-mass haloes, the distribution of SFRs is wider and thus allows haloes with higher than usual SFR to have high $f_\text{esc}$. On the other hand, the distribution of SFR in high-mass regions is relatively narrow and a typical massive halo does not have sufficient SFR to produce high $f_\text{esc}$ in our simulation. In the middle panel, $\Sigma_\text{SFR}$ shows a similar trend as SFR. In the bottom panel, the histograms are seen to be divided into two regions by $\overline{\Sigma}_\text{SFR}$, suggesting that this quantity is a potential discriminator for $f_\text{esc}$.}
    \label{fig:sfr_ssfr_mhalo}
\end{figure*}

\subsection{Median escape fractions}
To better connect the halo properties with these sources, in Fig.~\ref{fig:f_esc_no_dust} we show the dependence of the median halo intrinsic (dashed) and escaped (solid) photon rate ($\dot{N}$; top) and the median (solid) and mean (dashed) escape fraction ($f_\text{esc}$; bottom) on halo mass ($M_\text{halo} \in 10^8$--$10^{12}\,\Msun$; left), stellar mass ($M_\text{stars} \in 10^6$--$10^{10}\,\Msun$; middle), and star formation rate ($\text{SFR} \in 10^{-3}$--$10^2\,\Msun\,\text{yr}^{-1}$; right) for all selected haloes in different redshift ranges from $z = 5.5$--$13$. The grayscale backgrounds show the cumulative completeness of the halo selection, and the solid black lines indicate the mass of $90\%$ completeness. The absent grayscale background in the middle panels is due to the high completeness for $M_\text{stars} > 10^6\,\Msun$ and reflects our selection criterion of at least one star in haloes. We note that the mean escape fractions here are different from the global averaged escape fractions presented in Fig.~\ref{fig:M_pdf_all}, where the intrinsic luminosity of individual haloes are taken into account instead of a naive mean of escape fractions from haloes in the same mass range. These two results converge when haloes in the same mass ranges have similar intrinsic photon rates. Since this is not generally true, the median/mean escape fractions here behave differently from the ones in Fig.~\ref{fig:M_pdf_all}. We postpone a closer analysis to Sec.~\ref{sec:halo-to-halo variations}, where we discuss the variation across haloes. However, it is important to note that mean statistics are biased to represent the presence of high values, while medians are sensitive to resolution and completeness. In addition, the global averages include both bright outliers and stars outside the virial radius, and thus are slightly different than the $f_\text{esc}$ we defined in Eqn.~(\ref{eqn:f_esc}). The haloes at the low-mass end have escape fractions close to the birth cloud value of $f_\text{esc}^\text{cloud} = 0.37$, indicating that almost all photons reach the IGM once they escape the birth cloud. This maximal escape behaviour of low-mass haloes also justifies our assumption of adopting this value for non-selected haloes in Section~\ref{sec:global_statistics}. As the halo mass increases, the median $f_\text{esc}$ drops significantly with a minimum around $M_\text{halo} \sim 10^9\,\Msun$. The decrease of $f_\text{esc}$ at the low-mass end agrees with the results from the above-mentioned studies, but in our case this is clearly also driven by resolution limitations and a selection bias towards haloes with higher stellar-to-halo mass ratios, which might lead to an overestimation of $f_\text{esc}$ at the low-mass end. The median $f_\text{esc}$ remains low with $f_\text{esc} \sim 0.01$ at $z \sim 6$ in intermediate-mass haloes with an observed gradual increase in $f_\text{esc}$ as halo-mass increases up to about $M_\text{halo} \sim 10^{11}\,\Msun$ at all redshifts. Finally, $f_\text{esc}$ has a steep increase for haloes above this mass, as can be seen at lower redshifts. This may result from ionizing channels created by other mechanisms, e.g. powerful feedback associated with massive galaxies. We will discuss the impact of dust at the high-mass end in Section~\ref{sec:dust}. Comparing the curves in different redshifts, we see $f_\text{esc}$ increases with redshift in all mass ranges. Similar trends are also reported in several previous studies \citep{Eide2020, Faucher-Giguere2020, Ma2020}. Finally, comparing median and mean values of $f_\text{esc}$, we see an overall similar pattern but a much less extreme change in mean statistics below the mass of $90\%$ completeness. This implies the medians are more sensitive to the completeness compared to the means, although the two measures have reasonable agreement (or at least interpretations) in regions with full completeness.

Furthermore, since $f_\text{esc}$ is defined to be the ratio of the escaped and intrinsic photon rates, it is interesting to also examine the change in both quantities as a function of halo mass. At all redshifts, the medians of intrinsic and escaped photon rates increase with the halo mass, stellar mass, or star formation rate, although the curves become relatively flat at the low-mass end (Fig.~\ref{fig:f_esc_no_dust}). The MCRT procedure requires at least one star within the virial radius as a photon source in the calculation, and therefore the intrinsic photon rates do not further decrease as halo mass decreases at the low-mass end due to the limited mass resolution in the simulation. The escaped photon rates reach the lowest point at $\approx 10^9\,\Msun$, and reach the highest point at the high-mass end. In all redshifts, we notice higher photon rates at higher redshifts for a given halo mass range. This can be attributed to changes in the stellar population as galaxies assemble and evolve. In fact, the specific SFR in the halo decreases with time \citep[see Figure 1 of][]{KannanLIM2022}, which explains why this evolution is not apparent in the rightmost panel of Fig.~\ref{fig:f_esc_no_dust} showing the intrinsic photon rate as a function of the SFR. 
In fact, intrinsic photon rates are expected to be proportional to SFRs if galaxies have continuous star formation histories \citep{Kennicutt1998, Dijkstra2019}. The constant of proportionality depends on the IMF, and in our case we find $\dot{N}_\text{int} = 2.34 \times 10^{53}\,\text{s}^{-1}\,[\text{SFR}/(\Msun\,\text{yr}^{-1})]$ by fitting the values above the $90\%$ completeness with a linear function with a goodness of fit of $R^2 > 0.99$. The dip below $\text{SFR} \sim 0.1\,\text{Msun\,yr}^{-1}$ is a result of poorly resolved stochastic star formation histories in low-mass galaxies, which is exaggerated by plotting the median rather than the mean and in showing the gas-based instantaneous SFR against the star-based LyC production rate that is sensitive to the sampled ages.

The non-monotonic dependence of $f_\text{esc}$ on the halo mass strongly suggests that there are additional factors that can affect the escape of ionizing photons. In particular, the stellar mass and star formation rate are two of the more intuitive quantities since star-forming galaxies are believed to play a crucial role in reionization and many reported measurements of $f_\text{esc}$ come from starburst galaxies. Overall, the median curves in Fig.~\ref{fig:f_esc_no_dust} reveal a flatter or more moderate dependence on $M_\text{stars}$ and SFR, except at the highest ranges, while the redshift evolution is strong before the midpoint of reionization.

\begin{figure*}
    \centering
    \includegraphics[width=\textwidth]{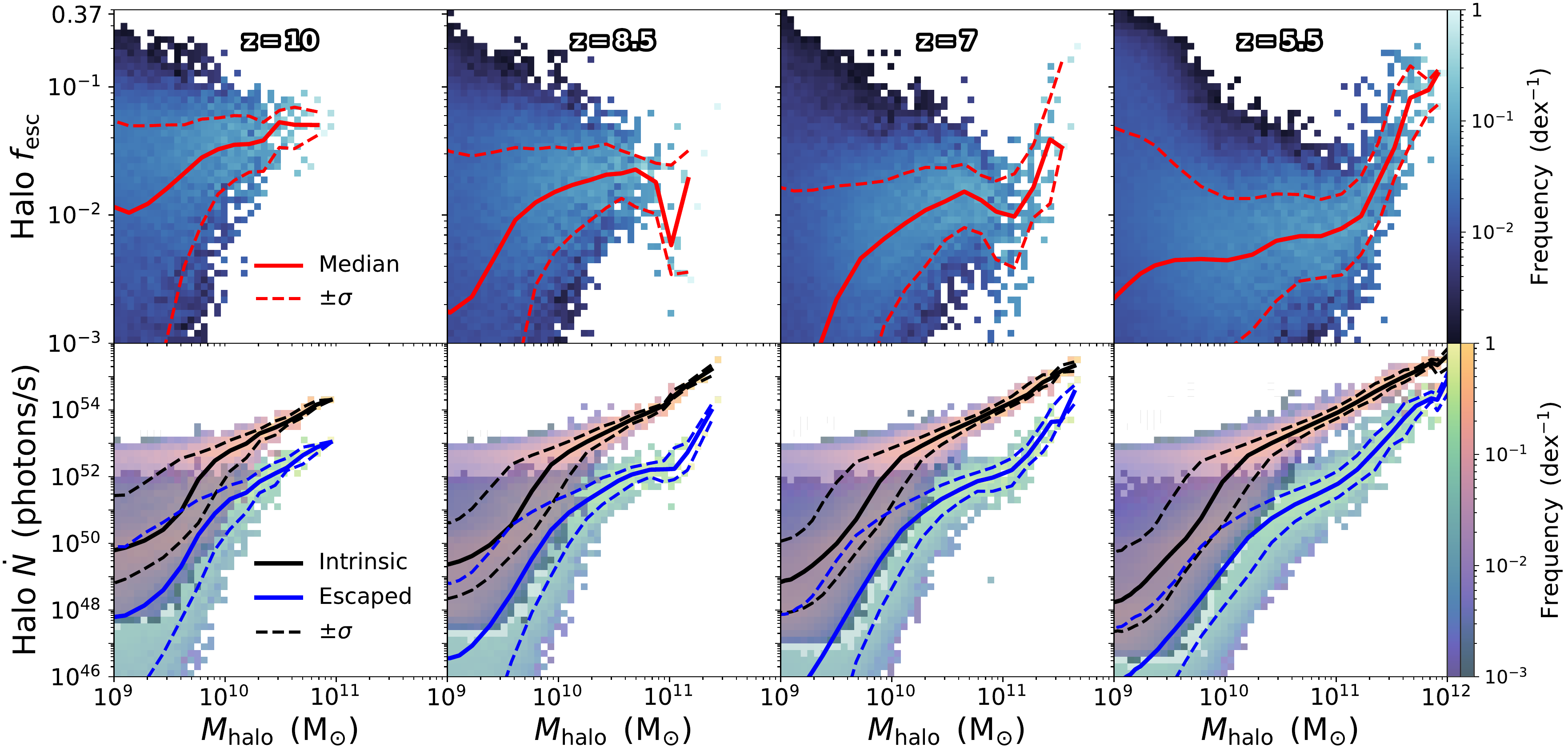}
    \caption{\textit{Top:} Distribution of halo escape fractions $f_\text{esc}$ as a function of halo mass $M_\text{halo}$ at redshifts of $z = \{10,8.5,7,5.5\}$, from left to right. Data below $10^9\,\Msun$ are not shown due to the lower completeness. The distribution is normalized by the number of haloes in each mass bin similarly to Fig.~\ref{fig:m_star_m200}. The red solid curves show the median $f_\text{esc}$ at a given mass bin, and the red dashed curves indicate the $16^\text{th}$ and $84^\text{th}$ percentiles. $f_\text{esc}$ starts with a wide distribution at the low-mass end and becomes narrower for more massive haloes ($M_\text{halo} \gtrsim 10^{10}\,\Msun$). At the massive end ($M_\text{halo} \gtrsim 10^{11}\,\Msun$), almost all haloes have $f_\text{esc} \sim 10\%$ although they only occur at $z \lesssim 7$. \textit{Bottom:} The distributions of intrinsic and escaped photon rates $\dot{N}$ as functions of halo mass. In both cases, the scatter is large at the low-mass end and becomes smaller above $M_\text{halo} \gtrsim 10^{10}\,\Msun$, with the escaped photon rate distributions being wider than the intrinsic ones.}
    \label{fig:hist2d_fesc_Mhalo_norm}
\end{figure*}

\subsection{Effective SFR surface density}
Beyond the cumulative and instantaneous star formation history, we expect the relative amount and density of absorbing hydrogen atoms in the surrounding gas reservoir to affect halo escape fractions. We therefore explore several of these features in Fig.~\ref{fig:sfr_ssfr_mhalo}. In the top panel we show the distribution of mean $f_\text{esc}$ in a 2D plane of SFR and $M_\text{halo}$ for galaxies at redshifts of $z = \{10,8.5,7,5.5\}$. The median and $1\sigma$ variation ($16^\text{th}$ and $84^\text{th}$ percentiles) of $f_\text{esc}$ at each mass bin is shown in the black solid and dashed curves, respectively. It is clear that the SFR increases as the halo mass increases in all mass ranges as expected. Galaxies with the highest $f_\text{esc}$ have a combination of lower mass and higher SFR as highlighted in the yellow regions. For slightly more massive haloes ($M_\text{halo} > 10^{10}\,\Msun$), $f_\text{esc} < 10\%$ even with high SFRs. In the low-mass region, the distribution of SFR at a given halo mass is wide and thus some haloes with SFRs $1$--$2$ orders of magnitude higher than the median can have high $f_\text{esc}$, which is expected due to starburst duty cycles. On the other hand, the distribution of SFR in the high-mass region is relatively narrow and most massive haloes in our simulation do not have SFRs far from the median to reach high $f_\text{esc}$. In addition, we note that since maximal escape fractions are achieved at $M_\text{halo} \sim 10^8\,\Msun$ due to the resolution limit and the mean $f_\text{esc}$ is higher than the median $f_\text{esc}$ at $M_\text{halo} \lesssim 10^{10}\,\Msun$, it is expected that most of the high escape fraction regions (yellow to orange) are located in this low-to-intermediate mass range.

In the middle panel of Fig.~\ref{fig:sfr_ssfr_mhalo}, we show the SFR surface density $\Sigma_\text{SFR} \equiv \text{SFR}_{r_e} / (2\pi r_e^2)$, with $r_e$ denoting the half stellar mass radius as this is observationally accessible \citep[although see][]{Trujillo2020}, and $\text{SFR}_{r_e}$ to be the SFR within the half stellar mass radius. This is additionally motivated by the fact that many of the LyC leakers have higher than average $\Sigma_\text{SFR}$ \citep{Naidu2020}. Recent observational efforts have also been focused on the dependence of $f_\text{esc}$ on $\Sigma_\text{SFR}$. In particular, \citet{Pahl2022} points out that a large sample is required to identify the relation between $f_\text{esc}$ and $\Sigma_\text{SFR}$. Our figure shows that at a given halo mass, galaxies with higher $\Sigma_\text{SFR}$ have higher $f_\text{esc}$, but it does not solely determine $f_\text{esc}$ either as there is a significant degree of mixing between higher and lower $f_\text{esc}$ values.

In an attempt to find a discriminating quantity that correlates with $f_\text{esc}$ in a monotonic manner, we modify the concept of $\Sigma_\text{SFR}$ by dividing SFR by the gas mass $M_\text{gas}$ and replacing a stellar-based radius with the virial radius $R_{200}$. We denote this new heuristic but theoretically motivated quantity as $\overline{\Sigma}_\text{SFR}=\text{SFR}/M_\text{gas}/R_{200}^2$. This particular combination is chosen since the gas mass of a galaxy roughly correlates with the amount of neutral hydrogen needed to be ionized before LyC photons can escape to the IGM. The bottom panel in Fig.~\ref{fig:sfr_ssfr_mhalo} shows the distribution of $f_\text{esc}$ with respect to $\overline{\Sigma}_\text{SFR}$ and the halo mass. The median values $\overline{\Sigma}_\text{SFR}$ drop slightly as halo mass increase from $10^8$ to $10^9\,\Msun$, but do not change significantly as the halo mass increases further. Moreover, there is a strong colour gradient in the vertical direction and relatively little variation along the horizontal direction indicating that $\overline{\Sigma}_\text{SFR}$ correlates with $f_\text{esc}$ in a simpler way than using $M_\text{halo}$ or SFR alone. However, we note that the escape of ionizing photons is a complex process, and the construction of $\overline{\Sigma}_\text{SFR}$ does not include other potentially important mechanisms such as AGN feedback in massive haloes.

\section{Variations in the escape fractions}
\label{sec:variations}
In the previous section, we discussed relationships between various physical quantities and the \textit{median}/\textit{mean} escape fractions. While this provides an intuitive picture for the sources of reionization, it does not fully represent the rich statistics across the entire ensemble of galaxies. In particular, bright outliers with $\dot{N}_\text{esc}$ above the median can play more important roles in the reionization process. In this section, we further discuss the distribution of $f_\text{esc}$ and photon rates to bring a more complete picture of halo escape fractions.

\begin{figure*}
    \centering
	\includegraphics[width=\textwidth]{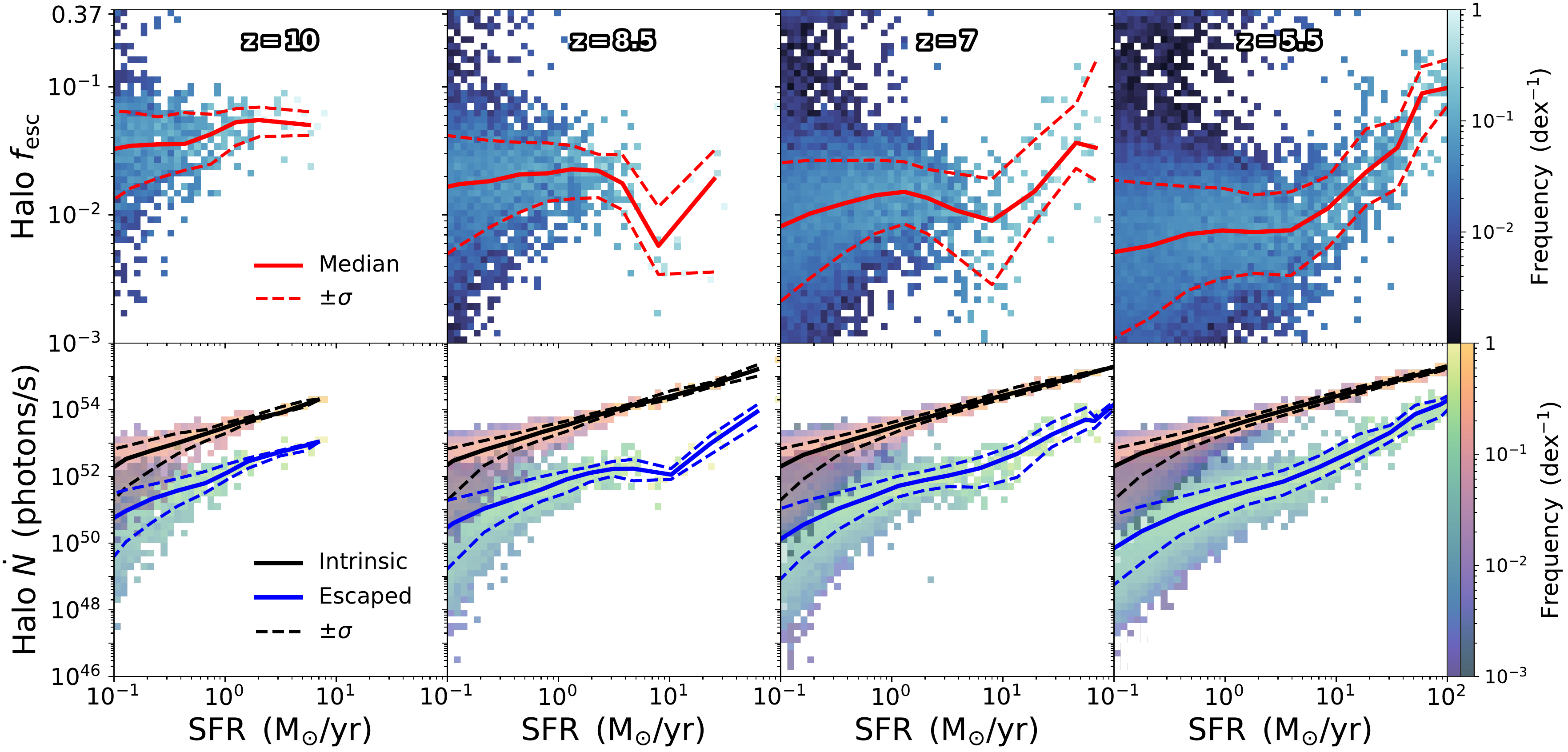}
    \caption{\textit{Top:} Distribution of halo escape fractions $f_\text{esc}$ as a function of the star-formation rate at redshifts of $z = \{10,8.5,7,5.5\}$, from left to right. All plotting styles and data representations are identical to Fig.~\ref{fig:hist2d_fesc_Mhalo_norm}, specifically the solid and dashed coloured curves indicate the median and 1$\sigma$ percentiles. As before, $f_\text{esc}$ have a large scatter at the low-SFR end and a narrower distribution as the SFR increases. \textit{Bottom:} The distributions of intrinsic and escaped photon rates $\dot{N}$ as functions of the SFR. Intrinsic photon rates correlate extremely well with SFR, such that the variation in escape fractions drives the IGM source distribution at a given SFR.}
    \label{fig:ndot_sfr_color_fesc}
\end{figure*}

\subsection{Halo-to-halo variations}
\label{sec:halo-to-halo variations}
In Fig.~\ref{fig:hist2d_fesc_Mhalo_norm} we show the normalized distribution of halo escape fractions $f_\text{esc}$ (top panels) along with intrinsic and escaped photon rates $\dot{N}$ (bottom panels) as functions of halo mass $M_\text{halo}$ at the specific redshifts $z = \{10,8.5,7,5.5\}$. The figure includes all selected haloes from the individual snapshots and each vertical histogram is normalized by the number of haloes in the given mass bin. The red solid lines in the top panels show the median halo $f_\text{esc}$ in each mass bin, and the dashed lines indicate the $16^\text{th}$ and $84^\text{th}$ percentiles. Likewise, the black (blue) solid lines show the medians of the intrinsic (escaped) photon rates, and the associated dashed lines correspond to the $16^\text{th}$ and $84^\text{th}$ percentiles. At low halo mass, $f_\text{esc}$ exhibits a wide distribution from $0.1$--$37\%$ with the medians between $1$--$10\%$ at $z = 10$ but extending down to $0.1\%$ at later redshifts. This is consistent with the picture in Fig.~\ref{fig:sfr_ssfr_mhalo} where $f_\text{esc}$ can vary for galaxies with similar masses because of different SFRs in different galaxies. In addition, a significant amount of medium-sized galaxies with masses below $\lesssim 10^{10}\,\Msun$ have $f_\text{esc} \lesssim 10^{-3}$  at $z = 7$. As $M_\text{halo}$ further increases, the distribution becomes narrower and most haloes have percent level escape fractions. The largest haloes with $M_\text{halo} \gtrsim 10^{11}\,\Msun$ generally have a higher $f_\text{esc}$, but are not commonly found in our simulated volume above redshift $z = 8.5$. In the bottom panel, both intrinsic and escaped photon rates again have wide distributions at the low-mass end and narrower distributions as the halo mass increases. This again agrees with the observation of wide SFR distributions at the low-mass end in Fig.~\ref{fig:sfr_ssfr_mhalo}. At the high-mass end, the intrinsic photon rates align well with $M_\text{halo}$, and the main contribution of the $f_\text{esc}$ comes from the variation in the escaped photon rates.

In Fig.~\ref{fig:ndot_sfr_color_fesc} we show the same distributions for $f_\text{esc}$ and $\dot{N}$ as functions of SFR ranging from $0.1$--$100\,\Msun/\text{yr}$ at redshifts $z = \{10,8.5,7,5.5\}$. Haloes with SFRs higher than $100\,\Msun/\text{yr}$ are rarely seen in the simulation, and the region below $0.1\,\Msun/\text{yr}$ is clipped due to the lower resolution below this range. In comparison to Fig.~\ref{fig:hist2d_fesc_Mhalo_norm}, SFRs better characterize the intrinsic photon rates of a halo as the distributions are generally narrower for SFRs than $M_\text{halo}$. At all redshifts, the variation in the intrinsic photon rates as a function of SFR is relatively small compared to that of halo mass, especially for haloes with SFRs higher than $1\,\Msun/\text{yr}$. This further implies that the wider distributions of the escaped photon rates are mostly from the complex escape fraction physics within the virial radius. That is, the strong sensitivity of $f_\text{esc}$ to ISM and CGM scale processes is the main factor in setting the variation in escaped sources into the IGM throughout the EoR.

Lastly, we argue that the wide distributions of intrinsic photon rates at the low-mass end leads to the difference between the global averaged escape fractions in Fig.~\ref{fig:M_pdf_all} and the mean $f_\text{esc}$ in Fig.~\ref{fig:f_esc_no_dust}. Since the brighter haloes usually have higher escape fractions as seen in Fig.~\ref{fig:sfr_ssfr_mhalo}, the global averaged escape fractions are higher than a naive mean. This is especially evident at $M_\text{halo} \lesssim 10^{10}\,\Msun$. At the high-mass end, the distributions of intrinsic photon rates are relatively narrow and the two escape fractions are very similar.

\begin{figure*}
    \centering
    \includegraphics[width=\textwidth]{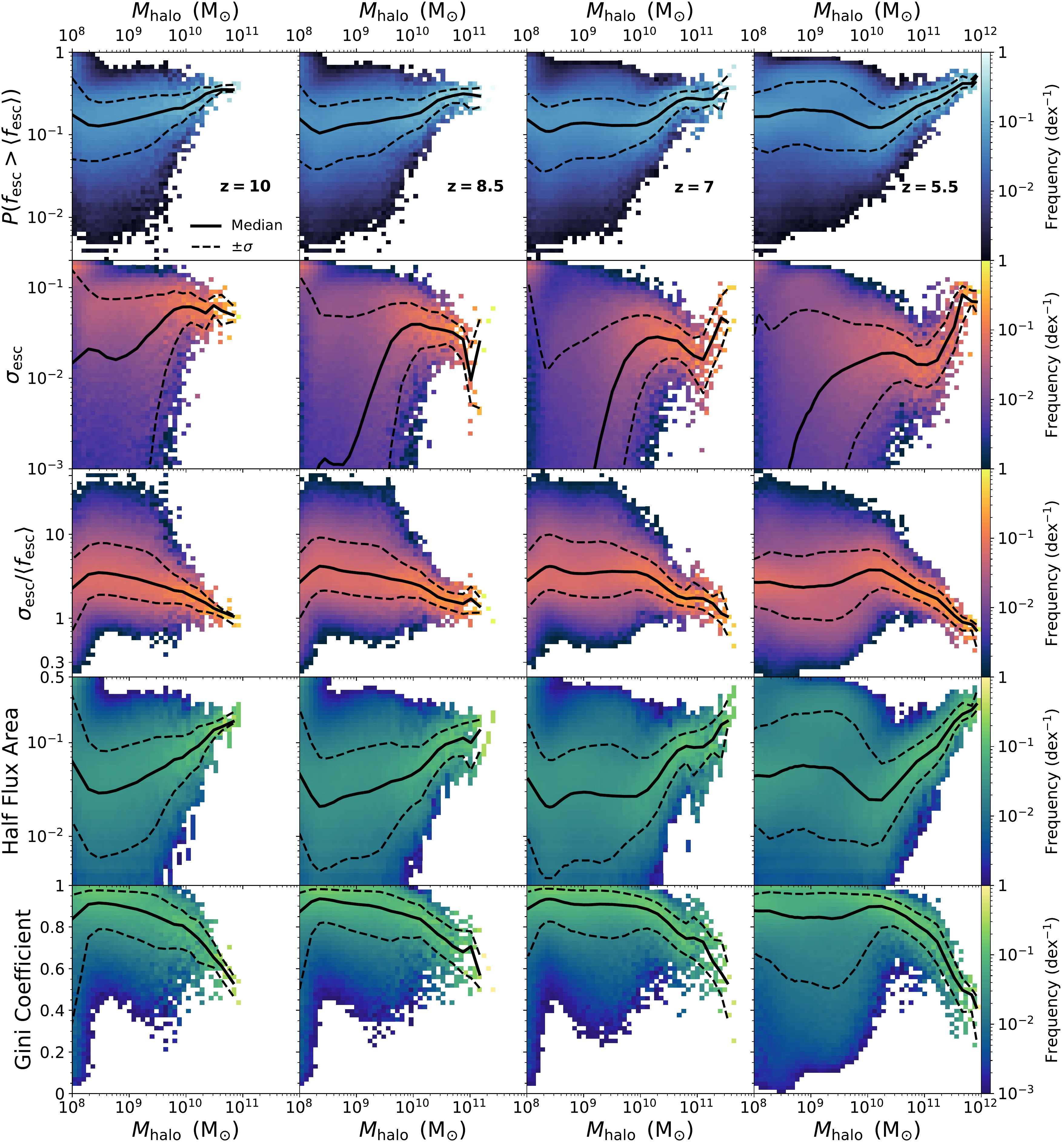}
    \caption{Distribution of covering fractions relative to the mean $P(f_\text{esc} > \langle f_\text{esc} \rangle)$ (top), standard deviations of escape fractions $\sigma_\text{esc}$ (second), and normalized standard deviations $\sigma_\text{esc} / f_\text{esc}$ (third), half flux area (fourth), and the Gini coefficient $G$ (bottom; defined in Eqn.~\ref{eqn:gini}) of all selected haloes as a function of $M_\text{halo}$ at redshifts of $z = \{10,8.5,7,5.5\}$. The distribution is normalized by the number of haloes in each mass bin similarly to Fig.~\ref{fig:m_star_m200}. The median and $16^\text{th}$ and $84^\text{th}$ percentiles are also shown in black solid and dashed lines, respectively. Most haloes have covering fractions $P$ falling between $0.05$ and $0.4$, i.e. a minority of sightlines influences the mean. There is no significant dependence on redshift but a qualitative increase for the highest mass haloes. The third panel shows that the standard deviations of sightline escape fractions in most haloes are above $\langle f_\text{esc} \rangle$, and the median of the normalized standard deviation decreases as halo mass and mean escape fraction increase. The fourth panel shows an increasing half flux area as halo mass increases, emphasizing the importance of sightlines that are not among the highest ones. The Gini coefficient in the bottom panel similarly shows a more isotropic and equitably distributed picture at the high-mass end.}
    \label{fig:cover}
\end{figure*}

\subsection{Sightline statistics}
In the previous sections, we have presented results from the angular-averaged $f_\text{esc}$ from all stellar sources escaping the surface of a sphere at $R_{200}$. However, it should not be surprising that the escape fractions also have a large sightline-to-sightline variability due to the anisotropic distribution of the matter within galaxies. In this subsection, we aim to capture the anisotropic structure using the ray-tracing techniques implemented in \colt to calculate $f_\text{esc}$ for observers oriented along each of $N=3072$ healpix directions of equal solid angle.

Analyzing the data from sightline statistics allows for deeper understandings into the underlying escape fraction physics. For example, we can answer the following questions: What is the variability of $f_\text{esc}$ from different sightlines? Do most of the photons escape through a handful of ionizing channels consistent with the concept of a covering fraction, or is a more isotropic picture favoured? What is the percentage of sightlines with $f_\text{esc}$ higher than the mean escape fraction? How do these answers also depend on halo properties?

To understand the anisotropic features of sightline dependent escape fractions in a more quantitative way, we define the following five quantities for each selected halo: (1) normalized covering fraction $P(f_\text{esc} > \langle f_\text{esc} \rangle)$ as the fraction of sightlines that have escape fractions higher than the mean escape fraction $\langle f_\text{esc} \rangle \equiv \frac{1}{N} \Sigma_{i=1}^N f_\text{esc,i}$, (2) standard deviation of the sightline escape fractions $\sigma_\text{esc}^2 \equiv \frac{1}{N} \Sigma_{i=1}^N (f_\text{esc,i} - \langle f_\text{esc} \rangle)^2$, (3) normalized standard deviation of the sightline escape fractions as $\sigma_\text{esc} / \langle f_\text{esc} \rangle$, (4) half flux area, which is defined as the fractional area of the sphere contributing half of the escaped photons when sorted from brightest to faintest, and (5) the Gini coefficient $G$ measuring the statistical dispersion in terms of equity of sightline distributions. Specifically, the Gini coefficient
\begin{equation}
    \label{eqn:gini}
    G = \frac{2\Sigma_{i=1}^N i f_\text{esc,i}}{N\Sigma_{i=1}^N f_\text{esc,i}} - \frac{N+1}{N} \, ,
\end{equation}
always lies in the range $[0,1]$ with values near unity representing significant anisotropy (escape dominated by a few sightlines) and small values representing significant isotropy (all sightlines contributing equally). As we will also make our escape fraction catalogues publicly available we defer more specialized viewing dependence analyses to future studies.

\begin{figure*}
    \centering
	\includegraphics[width=\textwidth]{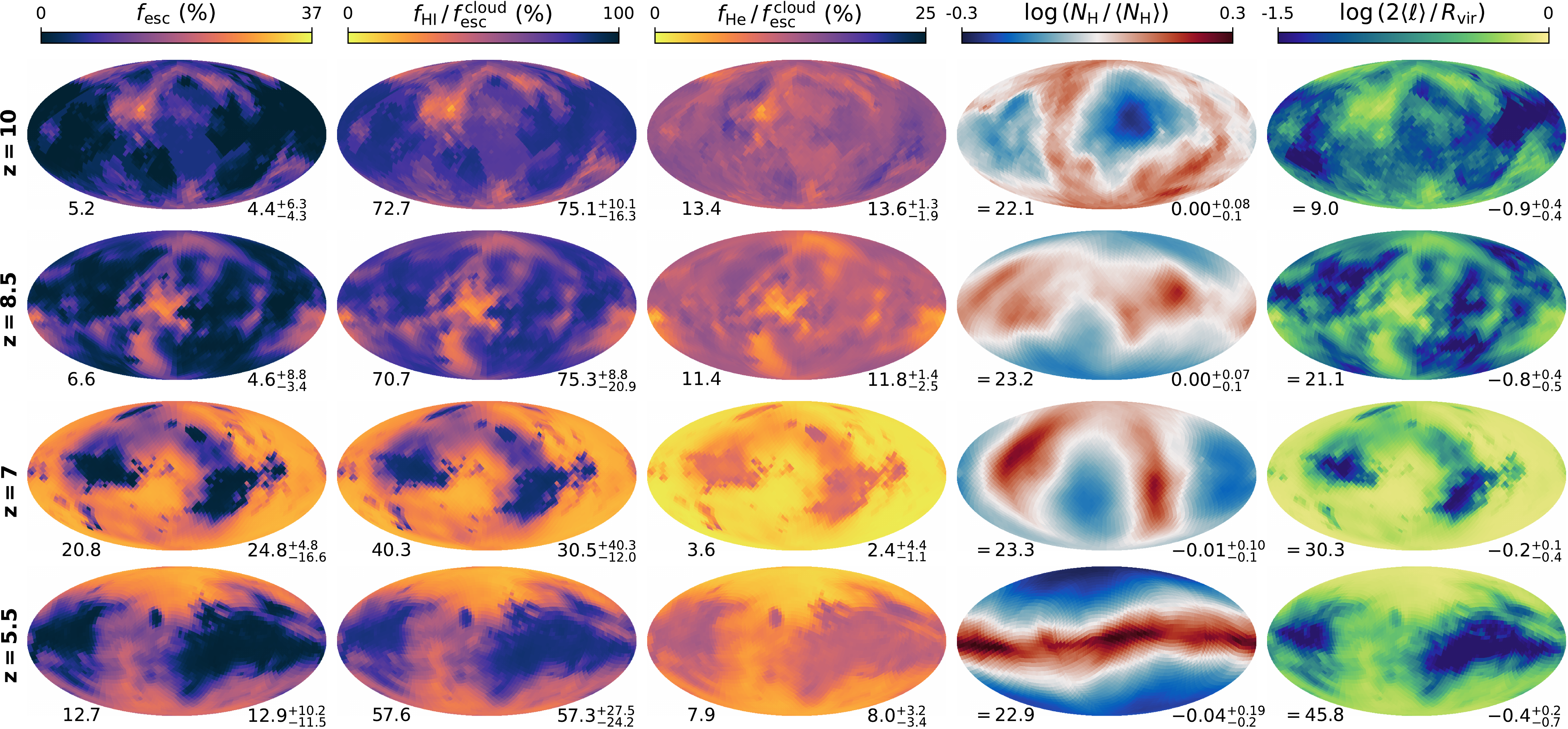}
    \caption{Healpix plots of various physical quantities in the most massive galaxies at redshifts of $z = \{10, 8.5, 7, 5.5\}$, respectively. From left to right these are: LyC escape fraction $f_\text{esc}$, hydrogen absorption fraction $f_{\HI}$ (normalized by $f_\text{esc}^\text{cloud}$), helium absorption fraction $f_{\HeI} + f_{\HeII}$ (also normalized), hydrogen column density $N_\text{H}$ (normalized by the mean hydrogen column density of 3072 directions), and mean absorption distance $\langle \ell \rangle$. The dust density has been set to zero in this calculation. In the lower left (right) of each map we provide the average (median and 1$\sigma$ variation) of all 3072 sightlines, with the exception of the hydrogen column density and absorption distance where for convenience we report the halo-dependent mean logarithmic column density (in $\text{cm}^{-2}$) and virial radius (in $\text{kpc}$), respectively. Formulae used in the calculation of absorption and mean absorption distance are discussed in Sec.~\ref{sec:methods}.}
    \label{fig:healpix_no_dust}
\end{figure*}

In Fig.~\ref{fig:cover}, we present the distributions of these five quantities as functions of halo mass at redshifts of $z = \{10,8.5,7,5.5\}$. The median and $1\sigma$ deviation ($16^\text{th}$ and $84^\text{th}$ percentiles) are also shown in black solid and dashed lines, respectively. In the top panel, $P$ represents the fraction of sightlines with higher-than-average escape fractions for each selected halo. At all redshifts, these normalized covering fractions have a wide distribution for less massive haloes $M_\text{halo} \lesssim 10^{10}\,\Msun$, ranging from $0.01$ to almost $1$ with $68\%$ of haloes having $P \in [0.05,0.4]$. The black solid curves suggest that in the median case, only $10$--$20\%$ of sightlines have $f_\text{esc}$ higher than their mean at all redshifts, revealing a substantial degree of anisotropy as $\langle f_\text{esc} \rangle$ is influenced by a small minority of sightlines. On the other hand, the distribution of $P$ is much narrower among high-mass haloes. For example, most of the haloes with $M_\text{halo}> 10^{11}\,\Msun$ have normalized covering fractions increasing to $30$--$50\%$ by $z=5.5$. This indicates that the sightline variability is more drastically different from halo to halo in low-mass haloes than in high-mass ones.

In the second and third panels, the standard deviation and normalized standard deviation provide a direct intuitive measure of the variability among sightlines. In particular, $\sigma_\text{esc} / \langle f_\text{esc} \rangle$ evaluate the standard deviations in the unit of the angle-averaged escape fractions. The third panel shows that at all redshifts the majority of haloes have sightline standard deviations larger than $\langle f_\text{esc}\rangle$. Similar to the distribution of the covering fractions $P$, standard deviations also have a wider distribution among lower mass haloes while transitioning to a smaller scatter for the highest mass haloes. Given these trends of increasing standard deviation and decreasing normalized standard deviation with halo mass, we interpret this as a transition from a low-escape high-anisotropic mode at $M_\text{halo} \lesssim 10^{10}\,\Msun$ to a more structured, large opening angle, and steady mode of escape at the highest masses.

The second to last panel in Fig.~\ref{fig:cover} shows the half flux area as defined above. The area gradually increases with halo mass, showing a notable increase above $M_\text{halo} \gtrsim 10^{10}\,\Msun$ at $z \lesssim 7$. The low median values for low-mass haloes indicate that the halo $f_\text{esc}$ are largely dominated by few sightlines with high escape fractions. In particular, half of the escaped photons are decided by the highest $1$--$10\%$ of sightlines for low-mass haloes, but the number rises to $\approx 25\%$ with less variation for the most massive haloes. This shows that the escape is not limited to a few high-$f_\text{esc}$ sightlines, and more ionizing channels are present in massive haloes. This picture provides a potential explanation of the increase in halo $f_\text{esc}$ with $M_\text{halo}$ as shown in Fig.~\ref{fig:f_esc_no_dust}. The more extreme intrinsic luminosities and structured environments of massive haloes foster the creation of either more ionized channels or a few dominant ones with larger opening angles, and thus leads to overall higher escape fractions.

Finally, the Gini coefficient $G$ in the bottom panel compares the contribution from high-$f_\text{esc}$ and low-$f_\text{esc}$ sightlines. At all redshifts, most low-mass haloes have $G$ close to unity, meaning the sightline distributions are highly skewed. $G$ deviates from unity as halo mass increases, and reaches $G \approx 0.5$ for the most massive haloes at $z = 5.5$. This shows a more equally distributed contribution from all sightlines.  Therefore, the measure from the Gini coefficient $G$ is consistent with the result from the half flux area.

\subsection{Angular structure correlations}
\label{sec:angular_distribution}
The anisotropic behaviour has also been discussed in other studies \citep[e.g.][]{Paardekooper2015, Trebitsch2017}, showing that most of the ionizing photons escape from galaxies through low column density channels. In Fig.~\ref{fig:healpix_no_dust} we provide visualizations of the angular distribution of the escape fraction $f_\text{esc}$, hydrogen absorption $f_{\HI}$, helium absorption $f_{\HeI} + f_{\HeII}$, hydrogen column density $N_\text{H}$ (normalized by the mean), and mean absorption distance $\langle \ell \rangle$ (normalized by the virial radius) in the most massive haloes at redshifts of $z = \{10,8.5,7,5.5\}$, respectively. Below each image, we show the mean (left) and median $\pm 1\sigma$ (right) statistics based on 3072 healpix directions, with the exception of the hydrogen column density and absorption distance where for convenience we report the halo-dependent mean logarithmic column density (in log $\text{cm}^{-2}$) and virial radius (in kpc) at the left. The absorption from hydrogen and helium is calculated using Eqn.~(\ref{eqn:f_esc_species}), with $f_\text{He}$ representing the combined absorption of \HeI\ and \HeII. The dust density has been set to zero and thus $f_\text{abs} = 0$ in this calculation. In general, by construction the sum of $f_\text{esc}$ and the absorption from different species equals the birth cloud escape fraction as can be calculated from the following equation: $f_\text{esc}^\text{cloud} = f_\text{esc} + f_{\HI} + f_\text{He} + f_\text{abs}$.

In greater detail, to more fairly compare the four different haloes, the hydrogen column density has been normalized by the angular-averaged value for each halo. Each sightline is the intrinsic photon rate weighted value from each star, i.e. $N_\text{H} \equiv \sum \dot{N}_{\text{int},i} N_\text{H,i} / \sum \dot{N}_{\text{int},i}$. Similarly, the mean absorption distance is normalized by the virial radius of the halo. The mean absorption distance corresponds to the typical distance photons travel before being absorbed, as defined in Eqn.~(\ref{eqn:mean_dist}). Note that the maximum mean optically thin absorption distance equals half the virial radius as can be seen by setting $k_a = 0$ in Eqn.~(\ref{eqn:mean_dist}). The median values in this example give $\langle \ell \rangle = \{0.57,1.56,10.34,8.32\}\, \text{kpc}$ with $R_\text{vir}=\{9.05,21.1,30.3,45.8\}\,\text{kpc}$ at $z = \{10,8.5,7,5.5\}$. Sightlines with particularly low absorption distances are expected to exist and are due to the geometry (e.g. covering) of neutral gas and star particle migration from dense environments.

Our results clearly demonstrate the anisotropic behaviour of $f_\text{esc}$. Moreover, the morphologies of the direct radiative transfer quantities (i.e. with some exceptions for $N_\text{H}$) are strikingly similar, indicating correlations among them at the sightline level. In particular, $f_\text{esc}$ strongly correlates with the mean absorption distance but anti-correlates with the hydrogen and helium absorption fractions. In other words, the photons can travel longer distances and escape the halo along low column density sightlines, resulting in lower $f_{\HI}, f_\text{He}$ and thus higher $f_\text{esc}$. Finally, disk-like structures from the forming galaxy manifest themselves in the bipolar distribution of the sightline statistics, with higher $f_\text{esc}$ in the direction perpendicular to the plane of the disk. The high anisotropy found in the values of $f_\text{esc}$ from various sightlines also implies the measured $f_\text{esc}$ can be highly dependent on the direction being measured (edge-on or face-on). It therefore presents a potential challenge for comparing simulations to observations, which are more or less randomly oriented. Of course, lower mass galaxies will have different morphologies and characterizing the anisotropic sourcing of LyC radiation during the EoR is worth a more detailed investigation, especially if small-to-large scale (ISM, CGM, and IGM) connections cascade to affect the bubble topology of reionization.

\section{Model uncertainties}
\label{sec:caveats}
We now discuss the main caveats of our analysis and when possible quantify the level of uncertainties, starting with dust absorption, AGN emission, and galaxy formation modelling. A discussion of resolution and convergence is presented in Appendix~\ref{sec:other}.

\begin{figure*}
    \centering
    \includegraphics[width=\textwidth]{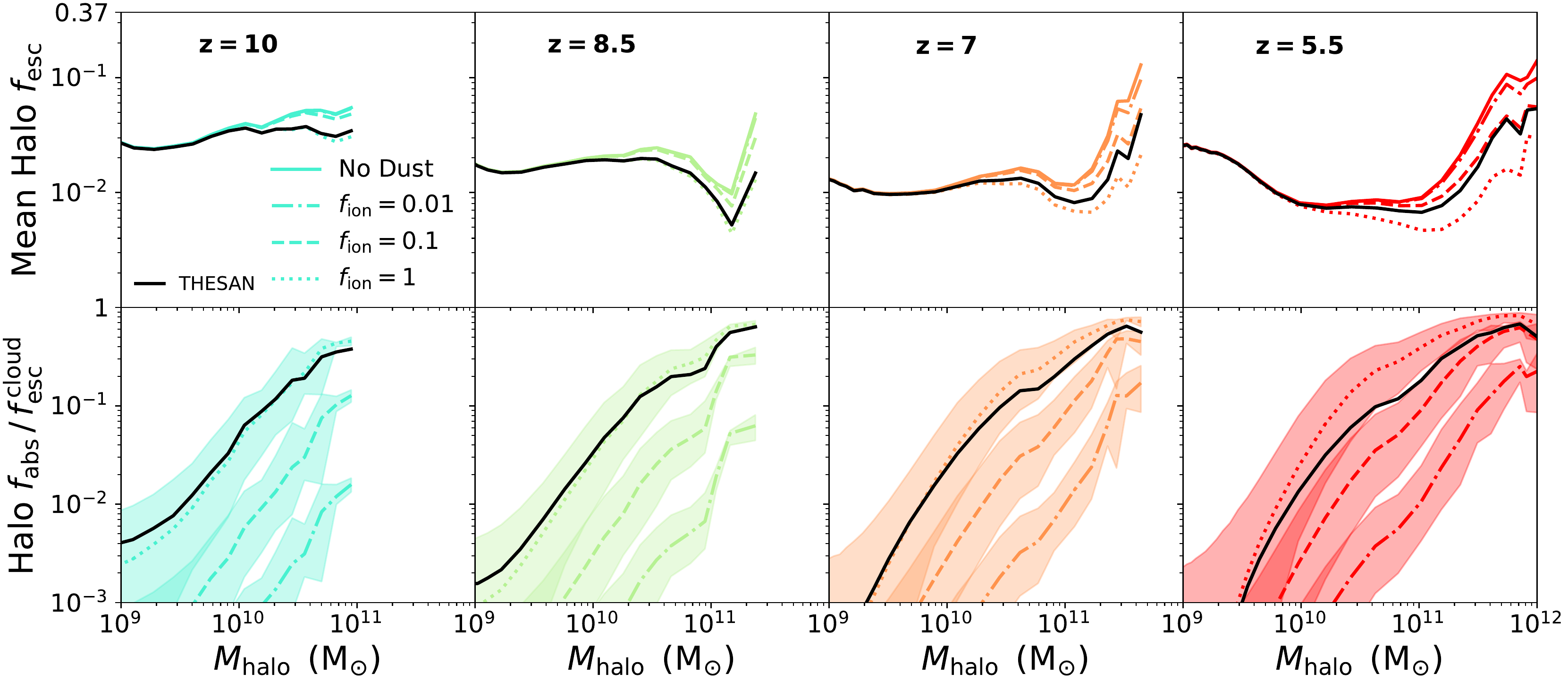}
    \caption{\textit{Top:} Mean values of halo escape fraction $f_\text{esc}$ as a function of $M_\text{halo}$ at $z = \{10,8.5,7,5.5\}$ for five different dust scenarios: no-dust (fiducial; solid), redshift-dependent dust-to-metal ratio from \citet{Vogelsberger2020} together with three different dust survival fractions in \HII regions $f_\text{ion} = \{0.01,0.1,1\}$ (dash-dotted, dashed, dotted), and on-the-fly dust modelling from \thesan (black solid). The inserted dust in all cases does not affect the overall dependence of $f_\text{esc}$ on the halo mass except at the high-mass end where a lower $f_\text{esc}$ results from the higher dust content. The on-the-fly simulated dust model lies between $f_\text{ion} = 0.1$--$1$. \textit{Bottom:} Median values of $f_\text{abs}$ as a function of halo mass shown for the same redshifts and same five dust scenarios as above. The shaded regions indicate the $16^\text{th}$ and $84^\text{th}$ percentiles across different haloes. $f_\text{abs}$ is calculated by Eqn.~(\ref{eqn:f_esc_species}) and increases as dust mass increases as expected. }
    \label{fig:dust}
\end{figure*}

\subsection{Dust absorption}
\label{sec:dust}
As briefly noted in Sec.~\ref{sec:methods}, \thesan does not include on-the-fly dust absorption for ionizing photons, mostly because dust modelling in the high-redshift Universe remains relatively unconstrained. Including dust would mainly translate to a slightly different behaviour for the radiation field in high-mass galaxies. While it is worth exploring the impact of on-the-fly dust absorption in \thesan, also in the context of subresolution ISM and feedback models, for consistency with the simulated ionization states we also do not include dust in our fiducial MCRT calculations. In this subsection, we focus on estimating the impact of dust on post-processed MCRT escape fractions.

Although \thesan incorporates the creation and destruction of dust, many physical processes relevant to dust occur at subparsec scales and thus the precise amount of dust in each cell might not be accurately determined. Thus, in addition to using the dust abundance modelled from the simulation, we also insert dust into the post-processing MCRT calculation using the redshift-dependent dust-to-metal ratio suggested by \citet{Vogelsberger2020} (model C), which has been calibrated to match observed UV luminosity functions. In the latter case, we also adopt three different dust survival fractions in the \HII regions of $f_\text{ion} = \{0.01,0.1,1\}$ to empirically explore scenarios with varying degrees of reduced dust abundances in ionized regions. This is a common parametrization used within the Lyman-alpha radiative transfer community to increase escape fractions with the justification of being a proxy for generally harsh environments \citep{Laursen2009}. Specifically, the destruction processes correlate with local feedback and unresolved turbulence or shocks, as well as other mechanisms such as radiation pressure \citep{Draine2011} or rotational disruption of dust grains by radiative torques in strong radiation fields \citep{Hoang2019}. We warn that values below $f_\text{ion} \lesssim 0.1$ may not be realistic \citep[see e.g.][]{Hu2019,Kannan2020}, however for our purposes the empirical model nonetheless provides a more continuous transition between the no dust and full dust scenarios. All the dust absorption opacities and anisotropic scattering are included based on the scattering albedo and Henyey--Greenstein phase function asymmetry parameter assuming the fiducial Milky Way dust model from \citet{Weingartner2001}.

In the top panel of Fig.~\ref{fig:dust} we show the mean halo $f_\text{esc}$ at redshifts of $z = \{10,8.5,7,5.5\}$ in the presence of dust with the models described above. Results using the simulation-determined dust distributions and constant dust-to-metal scaling are shown as black and coloured curves, respectively. For reference, the coloured solid curves show the values from the fiducial no-dust scenario, while the various dashed and dotted line styles indicate partial or full survival in \HII regions. The $f_\text{esc}$ dependence on $M_\text{halo}$ has a similar trend in all cases, with a noticeable difference at the high-mass end where $f_\text{esc}$ decreases as $f_\text{ion}$ increases. This is due to a higher amount of dust in high-mass haloes. In fact, the dust attenuation seems to compensate for the increased $f_\text{esc}$ to flatten out the high-mass end. The impact of dust attenuation is also higher at the lowest redshift ($z = 5.5$) due to the production of metals and dust throughout the stellar evolution process along with the increased presence of massive galaxies. We note that the abundance of simulation-modelled dust lies between $f_\text{ion}=0.1$ and $f_\text{ion}=1$ at all redshifts examined.

The bottom panel of Fig.~\ref{fig:dust} presents the median dust absorption $f_\text{abs}$ as calculated by Eqn.~(\ref{eqn:f_esc_species}) under the same five dust scenarios. The shaded regions illustrate the $1\sigma$ variation among the halo population. There is a clear increase in $f_\text{abs}$ with halo mass, again due to a larger amount of dust present in massive haloes, but it is also interesting to see the order-of-magnitude offsets with $f_\text{ion}$. The lowest redshift ($z = 5.5$) also reaches the most extreme values with $f_\text{abs} \approx \{0.25, 0.62, 0.83\}$ when $f_\text{ion} = \{0.01, 0.1, 1\}$. We emphasize that $f_\text{abs}$ represents the degree of dust pre-absorption and is therefore indicative of additional dust within ionized regions, which could shrink if on-the-fly dust absorption is included self-consistently. The impact on halo escape fractions is not as large as one might expect based on $f_\text{abs}$ alone because sightlines with low escape fractions will remain saturated. The global escaped photon rates are most affected if high $f_\text{esc}$ sightlines also undergo significant dust absorption. Keeping this in mind, our estimates are roughly consistent with other post-processing results from simulations with a multiphase ISM \citep[e.g.][]{Kado-Fong2020,smith2022disk,Tacchella2022}.

In addition to examining the effect of dust on the individual level, we also calculate the change in global escape fractions for different dust survival fractions as summarized in Table~\ref{tab:dust}. The dust effect is almost negligible at pre-reionization redshifts ($z \gtrsim 8$). In the most extreme case of full dust ($f_\text{ion}=1$), at the end of the simulation ($z = 5.5$) the global escape fraction is reduced to $\approx 57\%$ of the fiducial value from the no-dust scenario. We conclude that dust can be important during the latter half of the reionization process and incorporating on-the-fly prescriptions should be explored further in future simulation models. It remains unclear whether dust mainly adds degeneracy with other parameters such as the birth cloud escape fraction or if there are more fundamental differences.

\begin{table}
    \addtolength{\tabcolsep}{-2.5pt}
    \renewcommand{\arraystretch}{1.1}
    \begin{tabular}{@{} c cccccc @{}}
    \hline
    \multicolumn{7}{c}{Global ionizing escape fraction $f_\text{esc}\ \,[\%]$} \\
    \hline
    Redshift & No dust &  $f_\text{ion}=0.01$ & 
    $f_\text{ion}=0.1$ &  $f_\text{ion}=1$ & \thesan & $\frac{f_\text{ion}=1}{\text{No~dust}}$ \\
    \hline 
    13 & 14.3 & 14.3 & 14.3 & 14.3 & 14.3 & 99.8 \\
    11 & 10.7 & 10.7 & 10.7 & 10.7 & 10.7 & 99.5 \\
    10 & 10.9 & 10.9 & 10.9 & 10.8 & 10.8 & 99.2 \\
    9 & 7.91 & 7.91 & 7.89 & 7.83 & 7.82 & 99.0 \\
    8 & 6.37 & 6.35 & 6.30 & 6.16 & 6.18 & 96.8 \\
    7 & 5.82 & 5.69 & 5.45 & 5.12 & 5.27 & 88.0 \\
    6 & 5.92 & 5.63 & 4.96 & 4.22 & 4.67 & 71.3 \\
    5.5 & 4.89 & 4.40 & 3.51 & 2.77 & 3.31 & 56.7 \\ 
    \hline
    \vspace{.05cm}
    \end{tabular}
    \caption{The impact on the global escape fraction assuming the same dust scenarios as in Fig.~\ref{fig:dust}, including the fiducial no-dust case, redshift-dependent dust-to-metal ratio model with various empirical dust survival fractions of $f_\text{ion} = \{0.01,0.1,1\}$, and the \thesan-modelled dust. Note that here we only present the instantaneous global escape fractions at $z = \{13,11,10,9,8,7,6,5.5\}$, unlike in Fig.~\ref{fig:global_more} where the values have been smoothed for presentation purposes. Including dust into the post-processing MCRT calculations has little influence at higher redshifts due to the relatively low dust abundances and halo masses at pre-reionization epochs. The impact increases as redshift decreases with a factor of $\approx 2$ difference in the most extreme case by the end of the simulation.}
    \label{tab:dust}
\end{table}

\subsection{Active galactic nuclei}
The evident rise in $f_\text{esc}$ at the high-mass end as seen in Fig.~\ref{fig:f_esc_no_dust} suggests the potential presence of efficient ionizing mechanisms beyond the modest increase in stellar mass seen in Fig.~\ref{fig:m_star_m200}. In particular, AGN formed in massive haloes can create highly energetic jets and significant UV to X-ray radiation. In the context of \thesan, AGN also provide important thermal and kinetic feedback to the surrounding gas \citep{Weinberger2017}. These collectively help generate additional escape channels for ionizing photons. The effects of AGN on reionization have been discussed in some previous studies \citep[e.g.][]{Hassan2018,Trebitsch2018,Finkelstein2019,Dayal2020}, however \thesan offers a unique perspective because AGN sources are included on-the-fly throughout a large simulated volume. Thus, the effects of AGN on galaxies are included in our post-processing MCRT calculations in previous sections although only stars were considered as photon sources. In this subsection, we briefly present the results when treating AGN as additional photon sources.

\begin{figure}
    \centering
    \includegraphics[width=\columnwidth]{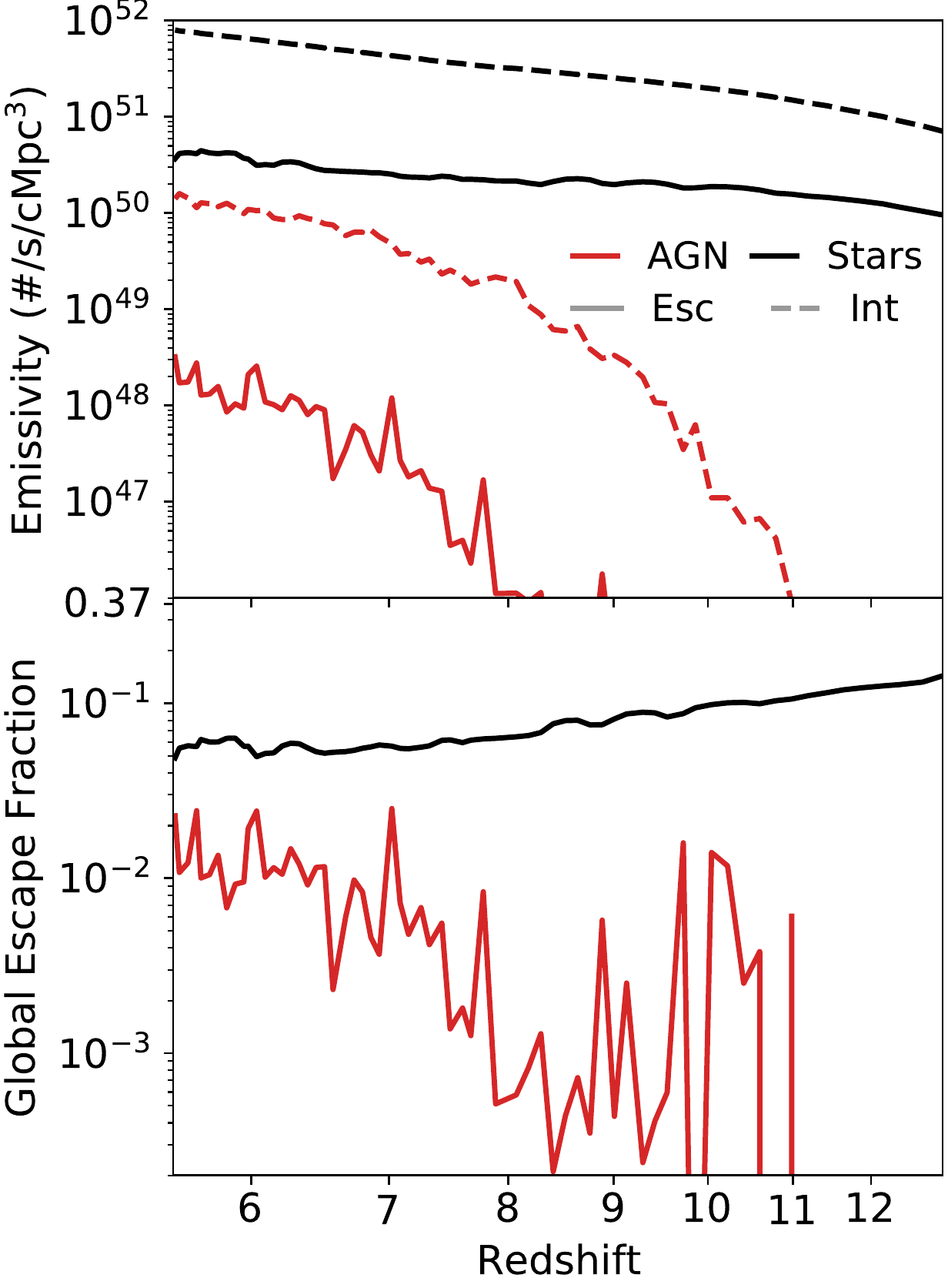}
    \caption{\textit{Top:} Global intrinsic (dashed) and escaped (solid) photon rate densities from AGN (red) and stars (black) as functions of redshift. The AGN curves are analogous to Fig.~\ref{fig:global_more} but with AGN as the only photon sources. Even at the lowest redshift ($z = 5.5$), AGN only contribute $1\%$ and $0.5\%$ of the total intrinsic and escaped photon rates, respectively. \textit{Bottom:} Comparison of the corresponding global escape fractions for AGN and stars as functions of redshift. AGN $f_\text{esc}$ are generally much lower than that of stars as a result of being located within galaxy centres. High redshift AGN statistics have large fluctuations due to the lack of massive galaxies in that regime given the limited simulation volume.}
    \label{fig:agn_z}
\end{figure}

In determining the AGN luminosity, we adopt the AGN spectral energy distribution (SED) from \citet{Lusso2015} with $35.5\%$ of bolometric luminosity (as given in the IllustrisTNG model) contributing to the LyC photons with a subgrid escape fraction of $100\%$. We then perform the same MCRT calculations with AGN as the only photon source, assuming isotropic emission and no dust absorption. In the top panel of Fig.~\ref{fig:agn_z} we show a comparison of the global intrinsic (escaped) photon rate density as dashed (solid) curves between AGN (red) and stars (black) as functions of redshift. The total intrinsic photon rate density $\rho_\text{int}$ from AGN is negligible at high redshifts, but rapidly rises as $z$ decreases due to the formation of more AGN within massive galaxies. By the end of the simulation at $z = 5.5$, $\rho_\text{int}$ remains below $2 \times 10^{50}\,\text{photons\,s}^{-1}\text{cMpc}^3$, which is approximately $100$ times lower than $\rho_\text{int}$ from stars. The low contribution from AGN is in agreement with \citet{Eide2020} and is also demonstrated in the original \thesan paper (Sec. 3.2.2 in \citet{Kannan2022}). Nevertheless, AGN could potentially increase or even dominate locally in simulations with substantially larger cosmic volumes or different physics modelling.

Similarly, the escaped photon rate density $\rho_\text{esc}$ from AGN is also negligible, and at $z = 5.5$ reaches global levels that are only about $0.5\%$ of the value from stars. In the bottom panel of Fig.~\ref{fig:agn_z} we show the corresponding global escape fractions showing that for AGN $f_\text{esc} \sim 1\%$, which is significantly below that of stars at all times. Contrary to the trend in the stellar $f_\text{esc}$, which drops slightly as redshift decreases, the AGN $f_\text{esc}$ rises as redshift decreases. The oscillating AGN $f_\text{esc}$ at high redshifts is due to the lack of statistics. We attribute the relatively low AGN $f_\text{esc}$ to the fact that AGN are positioned at the centres of galaxy gravitational potential walls, where the abundance of high-density gas results in higher self-shielding covering fractions. Thus, escaping from an AGN source requires substantially more ionizing photons to compensate for the systematically larger gas column densities. However, although AGN do not have a significant contribution to the global escaped photons, they potentially play important roles in creating \HII regions and ionizing channels for the more diffuse stellar sources.

\section{Conclusions}
\label{sec:summary}
In this paper, we have performed Monte Carlo radiative transfer calculations on all haloes from the \thesan simulation to study the production and escape of ionizing photons during the EoR. Our sample consists of haloes in the range $M_\text{halo} \in [10^8, 10^{12}]\,\Msun$, and covers $90\%$ of haloes with $M_\text{halo} > 1.2 \times 10^9\,\Msun$ at the midpoint of reionization at $z \approx 7.67$. The \thesan simulations have been shown to produce realistic galaxy and IGM properties that match the current observations and thus provide a self-consistent framework for studying the escape fractions of reionization-era galaxies \citep{Kannan2022,Garaldi2022,Smith2022}. The large volume of the simulation ($L_\text{box} = 95.5\,\text{cMpc}$) provides rich statistics for massive haloes ($M_\text{halo} > 10^{10}\,\Msun$) that are important in driving the reionization process but are often statistically underrepresented in previous studies. We summarize our main conclusions in the following points:

\begin{enumerate}
    \item Lower mass haloes ($M_\text{halo} \lesssim 10^9\,\Msun$) provide the majority of both intrinsic and escaped photons above $z \gtrsim 7$, while higher mass haloes ($M_\text{halo} \gtrsim 10^{10}\,\Msun$) dominate the photon budget thereafter. The most massive galaxies ($M_\text{stars} \gtrsim 10^9\,\Msun$) are particularly important at the later stages of reionization due to their proportionately high intrinsic luminosities. In fact, we find the decrease in the global escape fraction with decreasing redshift can be attributed to this transition from low-to-high mass dominance.
    \item Halo escape fractions have a non-trivial dependence on $M_\text{halo}$: $f_\text{esc}$ reaches a maximal value of $f_\text{esc}^\text{cloud} = 0.37$ for unresolved haloes ($M_\text{halo} \lesssim 10^8\,\Msun$), meaning that almost all ionizing photons escape after emerging from their local birth clouds. While this agrees with our physical intuition, it is essentially an artefact of numerical resolution in our model and we warn that previous studies reporting high $f_\text{esc}$ close to the resolution limit of their simulations also likely have this issue. Specifically, the median $f_\text{esc}$ decreases as halo mass increases, reaching a minimum once haloes are more robustly resolved around $3 \times 10^8\,\Msun$ at $z \lesssim 6$ and $10^9\,\Msun$ at $z \gtrsim 8$. After reaching the minimum value, $f_\text{esc}$ increases with halo mass, stellar mass, and SFR at all redshifts. This trend still holds even with the presence of dust attenuation, which has a larger effect on massive haloes. We note that while the halo mass corresponding to the minimum $f_\text{esc}$ shifts as the resolution changes, we find reasonable agreement in the global and resolved halo escape fractions with lower resolution simulations (see Appendix~\ref{sec:other}).
    \item Halo $f_\text{esc}$ are not a simple function of either $M_\text{halo}$, $M_\text{stars}$, SFR, or $\Sigma_\text{SFR}$. However, given a certain halo mass range, galaxies with higher SFR or $\Sigma_\text{SFR}$ tend to have higher $f_\text{esc}$. Furthermore, we found by considering SFR, gas mass, and the size of a halo, high $f_\text{esc}$ can be fairly well distinguished by a theoretical effective SFR surface density $\bar{\Sigma}_\text{SFR} = \text{SFR}/M_\text{gas}/R_{200}^2$.
    \item We find a wide range of values for $f_\text{esc}$ and $\dot{N}_\text{esc}$ at all redshifts, with lower mass haloes exhibiting broader distributions than high-mass ones. On the other hand, a tight relation can be observed between intrinsic photon rates and SFR, implying that the large variation of $f_\text{esc}$ is indeed due to the specific star formation histories and complex environmental dependence.
    \item Channels with lower escape fractions ($f_\text{esc}$) have higher hydrogen and helium absorption ($f_{\HI}$, $f_\text{He}$) and correspondingly lower absorption distances ($\langle \ell \rangle$). This implies that the break out of ionizing photons tends to be highly anisotropic. In the case of disk-like galaxies, escape fractions are shown to be higher in face-on directions and essentially zero for edge-on viewing angles. Lower mass galaxies tend to be more anisotropic in the sense that half of the escaping flux emerges from a smaller fraction of sightlines and the Gini coefficients are close to unity. The most massive galaxies exhibit a significant trend towards more isotropic and equitable escape.
    \item Including dust in the MCRT calculations has observable consequences for high-mass haloes, especially at low redshifts. By the end of the simulation at $z = 5.5$, the global escape fraction decreases from roughly $5\%$ in the no-dust scenario to nearly half that value with the added presence of dust for either the on-the-fly or constant dust-to-metal ratio models.
    \item We find that AGN intrinsic LyC production and escape fractions are generally lower than stellar sources. Therefore, while it is important to include AGN in large-volume simulations, they do not contribute significantly to hydrogen reionization.
\end{enumerate}

Our work analyzes millions of resolved galaxies in the large-volume \thesan simulation for haloes with mass between $10^8$--$10^{12}\,\Msun$. We demonstrate the dominant role of massive haloes in driving cosmic reionization by considering the escaped photons from the ensemble. We plan to include our escape fraction catalogues as part of the upcoming \thesan public data release, which can be utilized in follow-up analyses. Our results can also be used to inform and calibrate contemporary reionization models. In fact, we argue that including on-the-fly dust absorption is important for simulation volumes with significant numbers of massive haloes ($M_\text{halo} \gtrsim 10^{10}\,\Msun$). In the future, we plan to study the time-evolution of $f_\text{esc}$ from individual haloes by utilizing merger tree catalogues to correlate the production and escape of ionizing photons with properties of star formation and local environment in greater detail. Furthermore, the \thesan project will soon include high-resolution zoom-in resimulations of a wide range of galaxies including a multiphase ISM framework \citep{Marinacci2019,Kannan2020} and self-consistent meso-scale reionization environment inherited directly from the flagship simulation. These efforts will provide additional insights into the physics of escape from reionization-era galaxies and their evolved lower redshift counterparts for direct comparison to current and upcoming observational data.

\section*{Acknowledgements}
We thank the referee for comments and suggestions which have improved the quality of this work. JY acknowledges support from the MIT Undergraduate Research Opportunities Program. AS acknowledges support for Program number \textit{HST}-HF2-51421.001-A provided by NASA through a grant from the Space Telescope Science Institute, which is operated by the Association of Universities for Research in Astronomy, incorporated, under NASA contract NAS5-26555. MV acknowledges support through NASA ATP grants 16-ATP16-0167, 19-ATP19-0019, 19-ATP19-0020, 19-ATP19-0167, and NSF grants AST-1814053, AST-1814259, AST-1909831 and AST-2007355. The authors gratefully acknowledge the Gauss Centre for Supercomputing e.V. (\url{www.gauss-centre.eu}) for funding this project by providing computing time on the GCS Supercomputer SuperMUC-NG at Leibniz Supercomputing Centre (\url{www.lrz.de}). Additional computing resources were provided by the MIT Engaging cluster. We are thankful to the community developing and maintaining software packages extensively used in our work, namely: \texttt{matplotlib} \citep{matplotlib}, \texttt{numpy} \citep{numpy} and \texttt{scipy} \citep{scipy}.

\section*{Data Availability}
All \thesan simulation data will be made publicly available in the near future, including escape fraction catalogues. Data will be distributed via \url{www.thesan-project.com}. Before the public data release, data underlying this article will be shared on reasonable request to the corresponding author(s).



\bibliographystyle{mnras}
\bibliography{biblio}



\appendix

\section{Resolution dependence}
\label{sec:other}
\thesan also includes a suite of medium resolution simulations with different features aiming to investigate the impacts on reionization caused by different physics as described in \citet{Kannan2022}. In this appendix, we analyze the results from several of the medium resolution simulations to compare with the main results from the flagship \thesanone simulation explored throughout the body of this paper. We mainly focus on the \thesantwo simulation, which has the same initial conditions as \thesanone but with two (eight) times lower spatial (mass) resolution. When relevant we also include the \thesanwc simulation, where the birth cloud escape fraction is slightly higher to compensate for lower star formation in the medium resolution runs. In addition, \thesansdao assumes an alternative dark matter model that includes couplings to relativistic particles giving rise to strong Dark Acoustic Oscillations (sDAOs) cutting off the linear matter power spectrum at small scales, and \thesanhigh and \thesanlow are designed to better understand whether high or low-mass galaxies dictate the reionization process with only haloes above/below $10^{10}\,\Msun$ contributing to the escaped emission in \thesanhigh/\thesanlow. We note that in order to match the observed neutral hydrogen fraction, the different \thesan runs adopt different birth cloud escape fractions $f_\text{esc}^\text{cloud}$: \thesanone and \thesantwo have $f_\text{esc}^\text{cloud} = 0.37$, \thesanwc has $f_\text{esc}^\text{cloud} = 0.43$, \thesansdao has $f_\text{esc}^\text{cloud} = 0.55$, \thesanhigh has $f_\text{esc}^\text{cloud} = 0.8$, and \thesanlow has $f_\text{esc}^\text{cloud} = 0.95$. 

In addition to the medium resolution simulations, we also compare the results with a higher-resolution small box ($L_\text{box} = 12.5\, \text{cMpc}$, $N_\text{particles} = 2 \times 512^3$). This comes from a suite of simulations designed to investigate the impact of various reionization models on the properties of low-mass galaxies, and as such were simulated at much higher resolution than the original \thesan{} volumes. Our box, {\tt L8\_N512} denoted here as \thesanhr{}, has a baryonic mass resolution of $9.0 \times 10^4\,\Msun$, with a corresponding dark matter mass of $4.8 \times 10^5\,\Msun$, roughly a factor of 6.5 lower than \thesanone{}. The gravitational softening is scaled appropriately, with the gravitational forces softened on a scale of $0.85\,\text{ckpc}$, but all other parameters remain fixed with respect to \thesanone{}. This provides a unique opportunity to investigate the resolution dependence of the \thesan{} model, with these volumes discussed in detail in \citet{Borrow2022}.

\begin{figure}
    \centering
    \includegraphics[width=\columnwidth]{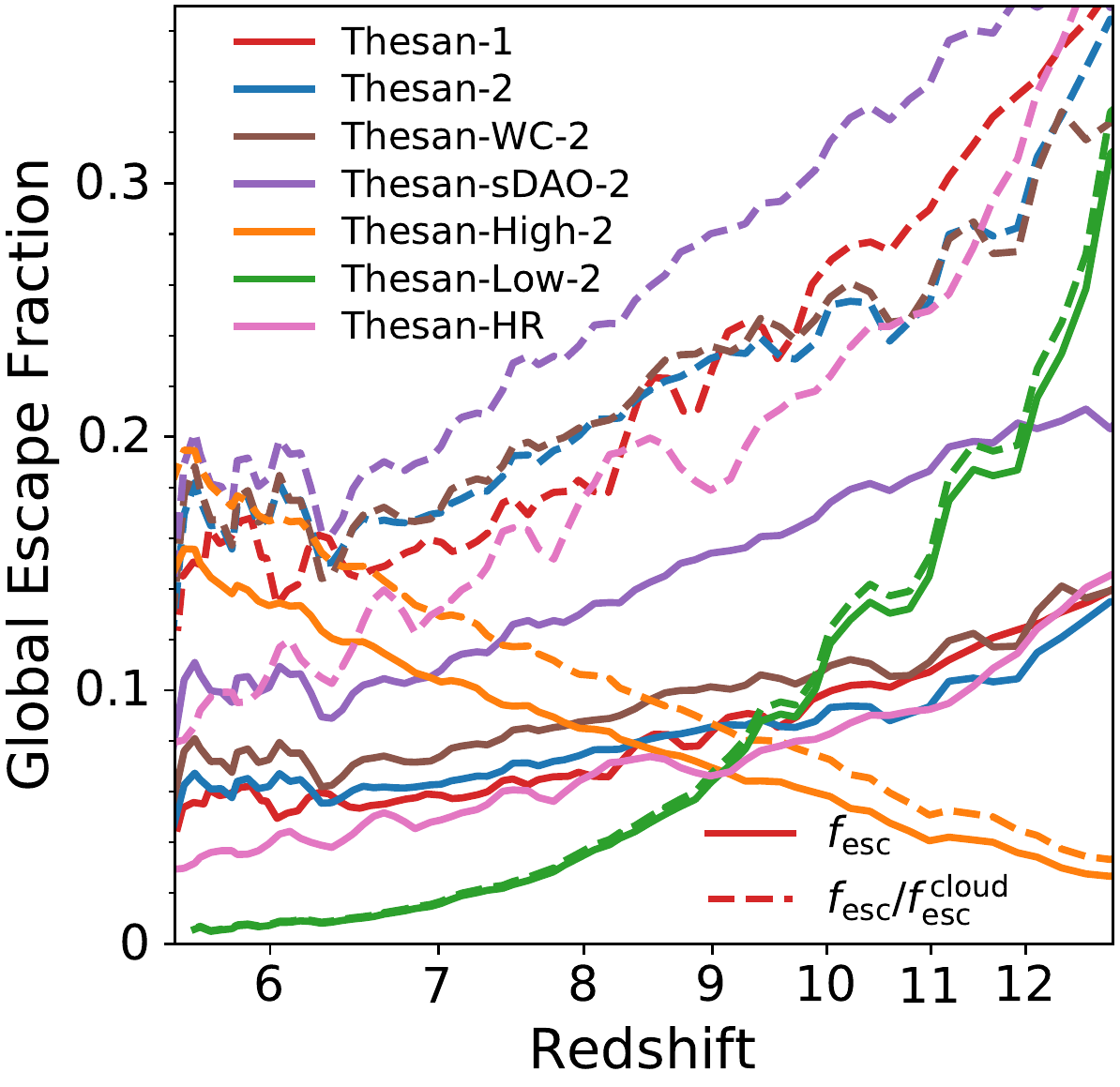}
    \caption{Comparison of the global photon rate weighted escape fractions from different \thesan simulations (see Fig.~\ref{fig:global_more}): \thesanone (red), \thesantwo (blue), \thesanwc (brown), \thesansdao (purple), \thesanhigh (orange), \thesanlow (green), and \thesanhr (pink). The original escape fractions are shown in solid lines, and the values after normalizing by the birth cloud escape fractions are in dashed lines. Results from \thesanone, \thesantwo, and \thesanwc agree reasonably well after the normalization, demonstrating the robustness of the global escape fractions. Discussions on the comparisons with other simulations can be found in the associated texts.}
    \label{fig:compare_sims_in_global}
\end{figure}

\begin{figure}
    \centering
    \includegraphics[width=\columnwidth]{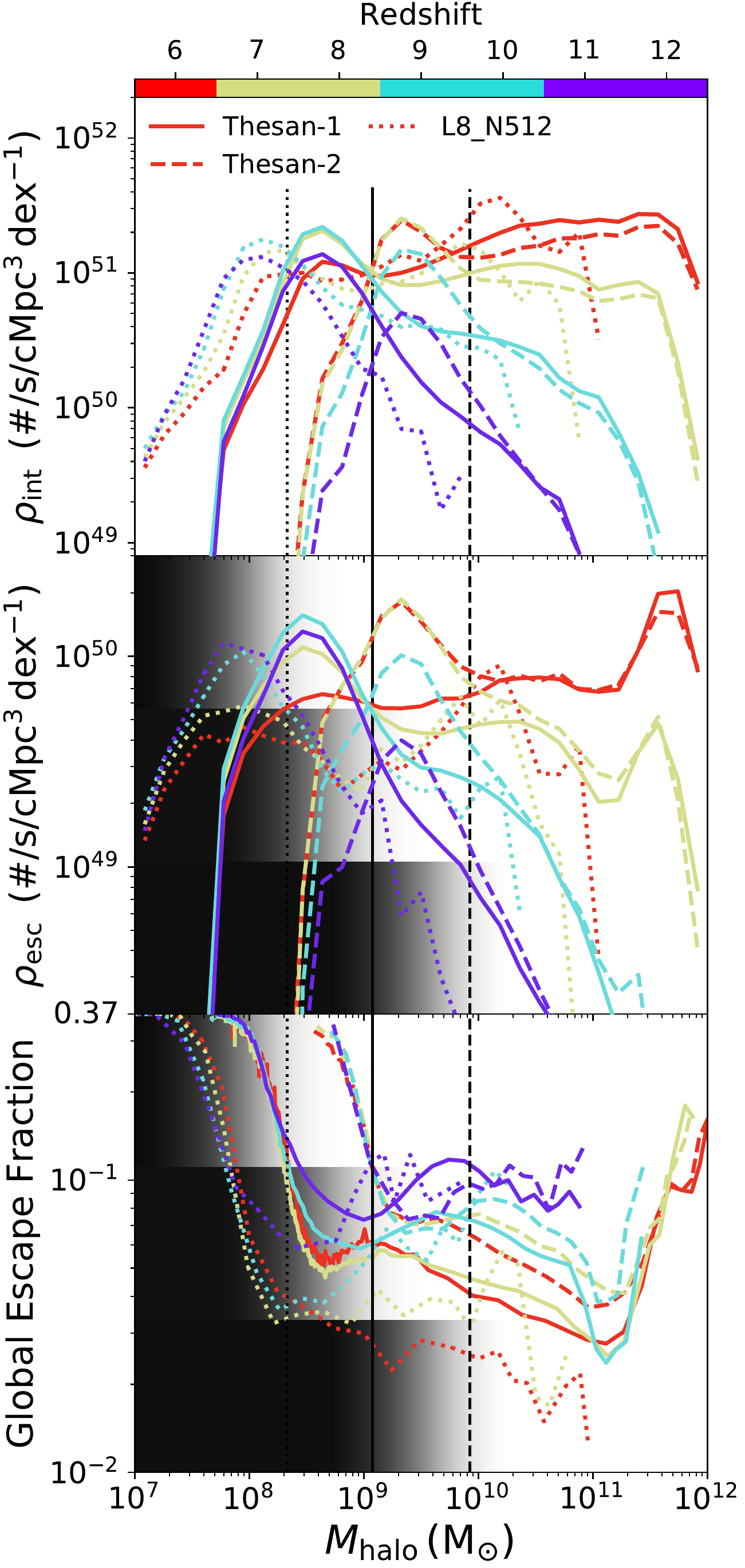}
    \caption{Resolution comparison between \thesanone (solid), \thesantwo (dashed), and \thesanhr (dotted) for the global intrinsic $\rho_\text{int}$ (top) and escaped $\rho_\text{esc}$ (middle) photon rate densities along with their escape fraction ratio (bottom) as functions of halo mass for several redshift ranges covering $z = 5.5$--$13$ (see Fig.~\ref{fig:M_pdf_all}). The grayscale backgrounds indicate the cumulative completeness, with vertical lines indicating the $90\%$ level. The intrinsic and escaped values in \thesanone and \thesantwo simulations agree at the high-mass end but differ at the low-mass end as halo structures and star formation histories become poorly resolved. \thesanhr shows a different structure at the high-mass end due to its limited volume. The peaks at the low-mass end shift to even lower masses in \thesanone and \thesanhr as the higher resolution allows the formation of lower mass galaxies. In the bottom panel, we see an overall reduction in global escape fractions as the resolution increases across most mass ranges, which is conveniently offset by the higher total intrinsic luminosities, such that there are similar amounts of escaped photons in all scenarios.}
    \label{fig:mhalo_combined_compare_sims}
\end{figure}

\subsection{Change in global statistics}
In Fig.~\ref{fig:compare_sims_in_global} we show the global escape fractions (solid) as a function of redshift for each of the above-mentioned \thesan variations. The \thesanone result is shown in the red curve for comparison (the same as in Fig.~\ref{fig:global_more}). The dashed lines show the values after normalizing the escape fractions by their birth cloud values to focus on differences in the MCRT calculations. The results from \thesanone, \thesantwo, and \thesanwc simulations are in reasonable agreement after the normalization, implying that the change in mass resolution has relatively little impact on the global result. Thus, the birth cloud escape fraction acts as a viable way to account for the reduced SFR density of the medium resolution runs. The trends in \thesanhigh and \thesanlow are also expected since the growing number of massive haloes that have a significant contribution to escaped emission only affect \thesanhigh, and are completely turned off in \thesanlow. The escape fractions from \thesanhr are seen to be lower at low-redshifts, which is likely due to the lack of high-mass haloes in the smaller box and consequently later reionization history rather than the improved resolution. Lastly, the ordering at $z \lesssim 7$ is as expected, i.e. if massive haloes start to dominate and are more or less converged with resolution then the curves should roughly line up near the end of the simulation.

Although the global results remain stable, adjusting the resolution creates a noticeable difference at the low-mass end as lower resolution limits the range of the least massive galaxies modelled. On the other hand, a simulation with a smaller box size limits the range of the most massive galaxies modelled and leads to differences at the high-mass end as well.  Similar to Fig.~\ref{fig:M_pdf_all}, Fig.~\ref{fig:mhalo_combined_compare_sims} shows the global intrinsic photon rate densities (top), escaped photon rate densities (middle), and escape fractions (bottom) from \thesanone (solid), \thesantwo (dashed), and \thesanhr (dotted). The three grayscale backgrounds indicate the cumulative completeness of \thesanhr, \thesanone, and \thesantwo from top to bottom with three vertical lines to indicate $90\%$ completeness. At the high-mass end, both intrinsic and escaped rates have similar values between \thesanone and in \thesantwo. Specifically, we note that they are reasonably converged for the dominant drivers of reionization ($M_\text{halo} \gtrsim 10^{10}\,\Msun$). The discrepancy between two simulations increases as the mass decreases and resolution artifacts become apparent. The positions of the peaks at the low-mass end shift to the left as the resolution increases and more low-mass galaxies are being resolved in \thesanone. The low-mass peak in \thesanhr shifts to further left due to its higher resolution. Furthermore, it shows a different structure at the high-mass end due its limited box size ($L_\text{box}=12.5\,\text{cMpc}$) compared to \thesanone and \thesantwo (both with $L_\text{box}=95.5\,\text{cMpc}$). The lack of massive haloes above $10^{11}\,\Msun$ is the main reason for the observed discrepancy. In terms of global escape fractions (bottom panel), the positions of the sharp upturn to maximal escape fractions are similarly set by the resolution. Quantitatively speaking, we found that increasing resolution leads to overall lowered values, which can be understood by requiring the total amount of photons to be the same for all simulations to have a consistent description of the reionization history. For example, the higher total luminosity in \thesanone due to a larger number of resolved stars is compensated by lower global escape fractions compared to \thesantwo.

Fig.~\ref{fig:mhalo_combined_compare_sims} shows that although the global emission is roughly converged at the high-mass end for simulations with the same box size, they generally do not agree at the low-mass end. However, in this case it is misleading to judge the convergence of a higher resolution simulation by comparing to a lower resolution one. It is better to compare to even higher resolution simulations, but to control computational cost it is inevitable to also reduce the box size when increasing resolution, which leads to different halo mass functions and other discrepancies. This being said, we are not aware of any reionization simulations claiming to have converged escape fractions. Systematic factors of a few are common, especially when considering individual haloes. For example, for the zoom-in simulations in \citet{Ma2020}, the absolute $f_\text{esc}$ does not converge in all mass ranges. This is also seen in other large-volume simulations such as \citet{Lewis2020} and \citet{Rosdahl2022}. We list several reasons that simulations might not demonstrate convergence: (i) changing the box size impacts the halo mass function especially at the high-mass end, and this leads to missing impacts on neighbouring haloes, (ii) some lower resolution simulations might not be able to capture key physical processes as escape fractions ultimately need to be resolved down to cloud scales, and (iii) some resolution dependent effects like feedback are difficult to account for in post-processing \citep{Trebitsch2018,Kimm2017}. To the best of our knowledge, there are no reionization simulations that can fully address the resolution convergence problem. However, we emphasize that in the above cases and ours that global results are sufficiently robust to maintain the same mechanisms for reionization and qualitative halo trends.

\begin{figure}
    \centering
    \includegraphics[width=\columnwidth]{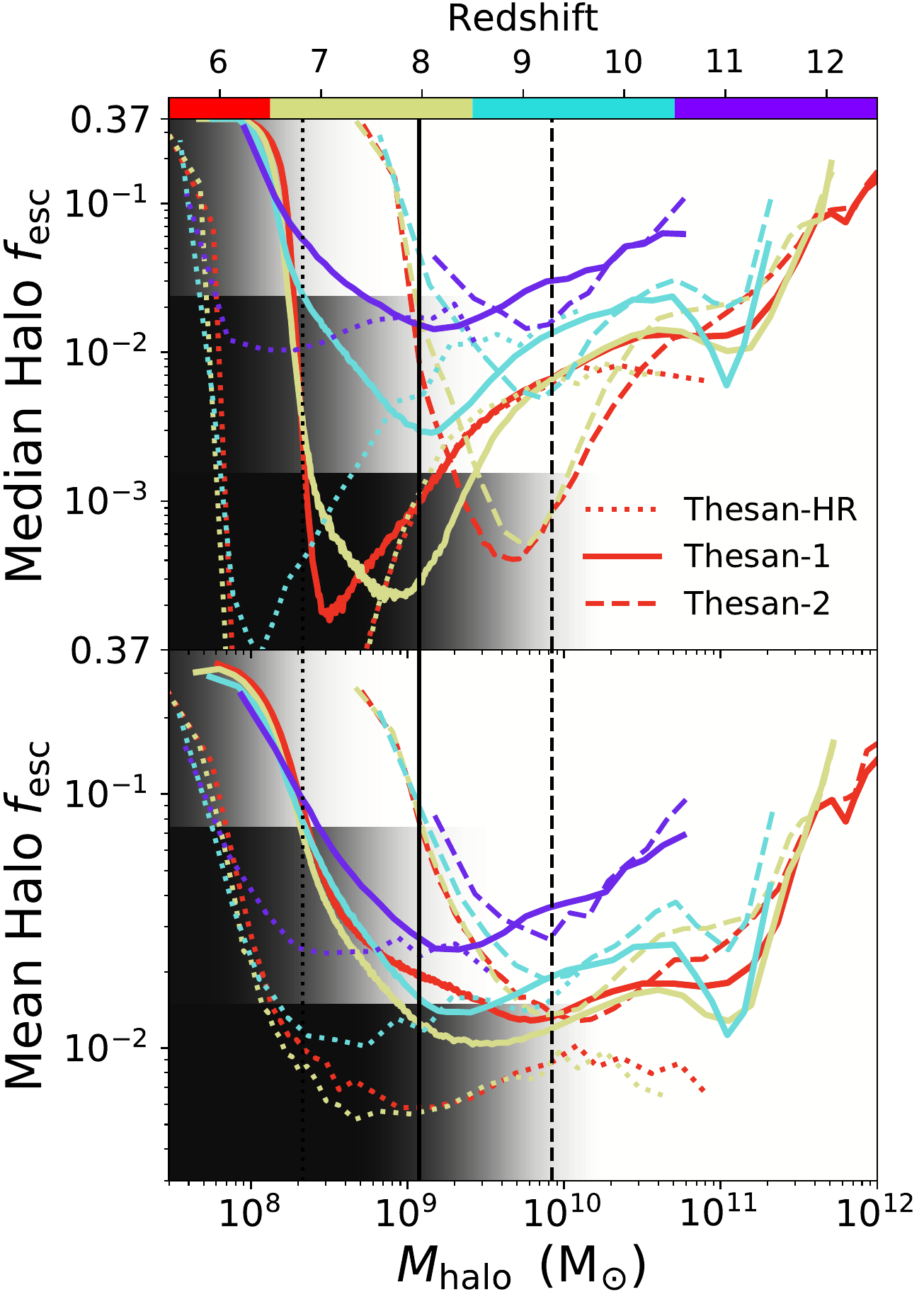}
    \caption{Resolution comparison between \thesanone and \thesantwo, and \thesanhr for the median (\textit{top}) and mean (\textit{bottom}) halo values of $f_\text{esc}$ as a function of $M_\text{halo}$ across several redshifts in the range $z = 5.5$--$13$ (see Fig.~\ref{fig:f_esc_no_dust}). The vertical lines correspond to $90\%$ completeness and the grayscale backgrounds indicate the cumulative completeness of \thesanhr, \thesanone, and \thesantwo from top to bottom. Three simulations admit the same pattern of the $f_\text{esc}$ dependence on halo mass as discussed in the text with a dip and gradual rise towards more massive haloes. The halo mass at which the minimal $f_\text{esc}$ occurs increases from $\approx 2 \times 10^8\,\Msun$ in \thesanhr to $\approx 10^9\,\Msun$ in \thesanone to $\approx 8 \times 10^9\,\Msun$ in \thesantwo as the resolution is lowered.}
    \label{fig:m200_fesc_compare_sims}
\end{figure}

\subsection{Change in individual halo dependence}
\label{appendix:halo_mass}
We also argue that the change in resolution does not significantly alter the pattern of the halo median/mean $f_\text{esc}$ dependence on halo mass. As shown in Fig.~\ref{fig:m200_fesc_compare_sims}, both the median (top) and mean (bottom) $f_\text{esc}$ dependence on $M_\text{halo}$ in \thesantwo and \thesanhr resembles the result from \thesanone except that the dip marking the transition from low completeness of low-mass haloes and sufficient resolution for richer substructures is shifted to higher masses for lower resolution simulations. This dip and gradual rise of $f_\text{esc}$ towards more massive haloes roughly corresponds to the point of $90\%$ completeness as indicated in the vertical lines, which is $\approx 2 \times 10^8\,\Msun$ in \thesanhr, $\approx 10^9\,\Msun$ in \thesanone, and $\approx 8 \times 10^9\,\Msun$ in \thesantwo. This is again due to resolution effects and this correspondence also implies that completeness can act as a reasonable proxy for the resolution. Overall, it is clear that the resolution requirements for obtaining robust halo escape fractions are likely more demanding than for other galaxy formation properties such as SFRs or gas fractions.

\section{Equation of State Cells}
\label{appendix:EoS}
\begin{figure}
    \centering
    \includegraphics[width=\columnwidth]{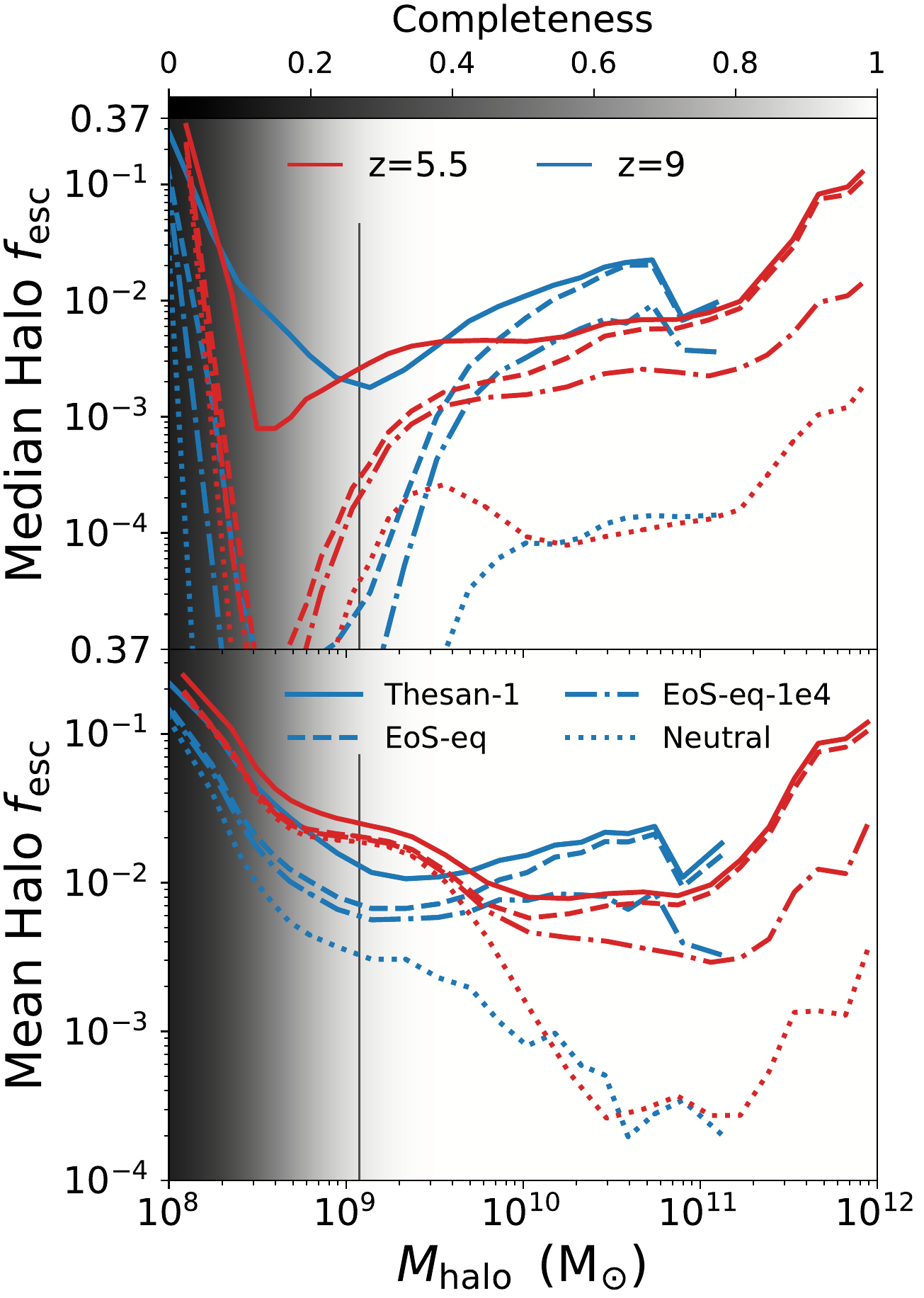}
    \caption{The change in median (top) and mean (bottom) $f_\text{esc}$ as a function of $M_\text{halo}$ while altering the ionization states in the equation of states (EoS) cells. We show the following four cases: (i) \thesanone (solid): using the ionization states from the simulation, (ii) EoS-eq (dashed): updating the EoS cells from converged values in photoionization equilibrium, (iii) EoS-eq-1e4 (dash-dotted): same as (ii) but first setting $T=10^4\,\text{K}$ in EoS cells, and (iv) Neutral (dotted): setting all EoS cells to be completely neutral.}
    \label{fig:EoS}
\end{figure}

In Sec.~\ref{sec:eos}, we have discussed the model uncertainty caused by EoS cells. To provide a more quantitative view, we briefly explore the impact of the on-the-fly coupling to the equation of state (EoS) for escape fractions by modifying the ionization states of EoS cells in post-processing. Fig.~\ref{fig:EoS} shows median (top) and mean (bottom) halo $f_\text{esc}$ as a function of halo mass at $z=5.5$ (red) and $z=9$ (blue) in the following four scenarios: (i) directly adopting the original ionization states from the \thesanone simulation as is done throughout  paper (\thesanone, solid), (ii) performing MCRT photoionization equilibrium calculations \citep[as in][]{smith2022disk} to update the EoS ionization states while retaining the states of non-EoS cells (EoS-eq, dashed), (iii) same as (ii) but in addition also requiring the temperature of EoS cells to be $T = 10^4\,\text{K}$ (EoS-eq-1e4, dash-dotted), and (iv) setting all EoS cells to be fully neutral (Neutral, dotted). The grayscale background is again shown to represent the cumulative completeness of the halo selection and the vertical lines show the $90\%$ level. We find the Neutral scenario shows a much lower $f_\text{esc}$ compared to the fiducial \thesanone results and establishes a lower limit on the impact of the EoS on $f_\text{esc}$. This comparison reflects that EoS cells are mostly ionized in simulations. To assess the validity of the effective ionization states in EoS cells, we compare \thesanone with EoS-eq, where photoionization equilibrium ionization states are used. From Fig.~\ref{fig:EoS}, we see that these two results agree well at the high-mass end in both median and mean. At the low-mass end, they show reasonable agreement in mean statistics but differ in median upon entering the range with low completeness. Overall, the escape of ionizing photons using ionization states from the simulation are comparable to those obtained by the photoionization equilibrium. Moreover, lowered values from EoS-eq-1e4 compared to EoS-eq imply that some ionizing photons escape as a result of collisional ionization from artificially high temperatures. Quantitatively, we find the global escape fraction at $z = 5.5\,(9)$ decreases from $4.89 \%\,(7.91\%)$ in the fiducial model to $4.30\%\,(7.42\%)$ in EoS-eq, $2.43\%\,(7.08\%)$ in EoS-eq-1e4, and $1.67\%\,(4.25\%)$ in Neutral, i.e. a modest factor $\lesssim 2$ uncertainty if the birth cloud escape fraction accounts for local cold phase absorption while most transport is through the hot ionized phase \citep[e.g.][]{Buck2022}. We conclude that on-the-fly EoS coupling should be studied in greater detail as the low $f_\text{esc}^\text{cloud}$ calibration is a direct result of the adopted EoS absorption model \citep[see also][]{Kostyuk2022}. The fiducial passive RT method used throughout the body of this paper is the correct approach for interpreting \thesan escape fractions because it most closely reflects the on-the-fly results for this particular model.

\bsp	
\label{lastpage}
\end{document}